\documentclass[12pt,preprint]{aastex}
\usepackage{amssymb}
\usepackage{pifont}

\usepackage{graphicx}
\RequirePackage{longtable}%

\shorttitle{Phase-Resolved Spectroscopy of Pulsars}
\shortauthors{M.Y. Ge et al.}

\begin{document}

\title{X-ray Phase-Resolved Spectroscopy of PSRs B0531+21, B1509-58, and B0540-69 with RXTE}

\author{M. Y. Ge, F. J. Lu, J. L. Qu, S. J. Zheng, Y. Chen, and D. W. Han}

\affil{Key Laboratory for Particle Astrophysics, Institute of High
Energy Physics, Chinese Academy of Sciences, Beijing 100049, P.R.
China; gemy@mail.ihep.ac.cn}

\begin{abstract}
The Rossi X-ray Timing Explorer ({\sl RXTE}) has made hundreds of
observations on three famous young pulsars (PSRs) B0531+21 (Crab),
B1509-58, and B0540-69. Using the archive {\sl RXTE} data, we have
studied the phase-resolved spectral properties of these pulsars in
details. The variation of the X-ray spectrum with phase of PSR
B0531+21 is confirmed here much more precisely and more details are
revealed than the previous studies: the spectrum softens from the
beginning of the first pulse, turns to harden right at the pulse
peak and becomes the hardest at the bottom of the bridge, softens
gradually until the second peak, and then softens rapidly. Different
from the previous studies, we found that the spectrum of PSR
B1509-58 is significantly harder in the center of the pulse, which
is also in contrast to that of PSR B0531+21. The variation of the
X-ray spectrum of PSR B0540-69 seems similar to that of PSR
B1509-58, but with a lower significance. Using the about 10 years of
data span, we also studied the real time evolution of the spectra of
these pulsars, and no significant evolution has been detected. We
have discussed about the constraints of these results on theoretical
models of pulsar X-ray emission.

\end{abstract}


\section{Introduction}

Phase resolved X-ray spectroscopy is very important for the
verification of pulsar emission models. Currently there are two
kinds of popular models to interpret the high energy emission of
pulsars: the polar gap model and the outer gap model. The polar gap
models assume that primary charged articles are accelerated above
the neutron star surface and the $\gamma$-rays result from a
curvature radiation and inverse Compton induced pair cascade in a
strong magnetic field \citep{Sturner and Dermer(1994),Daugherty and
Harding(1996)}. The radiated spectra from these cascades are hard
with photon indices 1.5-2.0 \citep{Harding and Daugherty(1998)}.
Outer gap models \citep{Cheng et al.(1986a),Cheng et al.(1986b),
Romani(1996)} assume that acceleration occurs along null charge
surfaces in the outer magnetosphere and that $\gamma$-rays result
from photon-photon pair-production-induced cascade. \cite{Cheng et
al.(2000)} (hereafter CRZ model) have re-considered the
three-dimensional magnetosphere gap model by introducing various
physical processes (including pair production, surface field
structure, and reflection of $e^{\pm}$ pairs) to determine the
three-dimensional geometry of the outer gap. \cite{Hirotani et
al.(2003)} pointed out that the large current in the outer gap can
change the boundary of the outer gap \citep{Hirotani and
Shibata(2001), Hirotani et al.(2003)}. The modified outer-gap model
could well predict the double peak profile and fit the phase-resolved X-ray
spectra of the Crab pulsar using the synchrotron self-Compton mechanism
\citep{Jia et al.(2007), Tang et al.(2008)}. \cite{Zhang and
Cheng(2000)} calculated the optical, X-ray and gamma-ray light
curves and spectra of  PSRs B0540-69 and  B1509-58 using the CRZ
model. They also derived the magnetic inclination angles, viewing
angles, and the thicknesses of the emission regions of these two
pulsars. High precision X-ray light curves and phase resolved X-ray
spectra of young pulsars could be used to constrain these models as
well as their physical parameters.

We choose three young and bright pulsars, PSRs B0531+21,
B1509-58, and B0540-69, to study their timing behaviors and the
phase-resolved spectra. Their periods are 33ms, 150ms, and 50ms, and
period derivatives 4.23$\times$ 10$^{-13}$ss$^{-1}$,
1.5$\times$10$^{-12}$ss$^{-1}$, 4.8$\times$10$^{-14}$ss$^{-1}$
respectively. They have similar characteristic spinning down
age($\sim$1.5 kyr), magnetic field(~$10^{12}$ G) and spin down power
($\sim10^{38}$ erg s$^{-1}$). From the timing results, their braking
indices are determined as 2.509$\pm$0.001 \citep{Lyne et al.(1988)},
$\sim$2.83 \citep{Kaspi et al.(1994), Simon Johnston and David
Galloway (1999), Livingstone et al.(2011)} and $\sim$2.10
\citep{Zhang et al.(2001),Cusumano
et al.(2003), Livingstone et al.(2005)}. The pulsed X-ray
luminosities of Crab and PSR B0540-69 are 1.3$\times$10$^{36}$\,erg
s$^{-1}$ , 4.0$\times$10$^{36}$\,erg s$^{-1}$ \citep{Kaaret et
al.(2001)}, and that for PSR B1509-58 is 4.7$\times$10$^{35}$\,erg
s$^{-1}$ \citep{Marsden et al.(1997)}. Their detailed
characteristics are listed in Table. \ref{table:pulsarpar}.

The X-ray spectra of these three pulsars can be well fitted by
power-law model with photon indices of 2.022 \citep{Kuiper et
al.(2001)}, 1.358 \citep{Marsden et al.(1997)} and $\sim$1.9
\citep{Finley(1993), Mineo et al.(1999), Kaaret et al.(2001), Plaa
et al.(2003), Campana et al.(2008)} respectively. However, the
photon indices change at different phase, especially for the Crab
pulsar. The variation of X-ray spectrum of the Crab pulsar with
pulse phase was first measured by \cite{Toor and Seward(1977)} and
then confirmed by \cite{Pravdo and Serlemitsos (1981)}. With the
Rossi X-ray Timing Explorer ({\sl RXTE}) and Beppo-SAX observations,
the phase-resolved X-ray spectrum of the Crab pulsar was studied and
it was found that the spectrum softens starting at the leading edge
of the first peak until its intensity maximum of the first peak; the
spectrum hardens in the interpeak region; and the spectrum softens
throughout the second peak \citep{Pravdo et al.(1997), Massaro et
al.(2000), Massaro et al.(2006)}. \cite{Rots et al.(1998)} obtained
the phase resolved X-ray spectra of PSR B1509-58 with  {\sl RXTE}
observations and suggested that all spectra are consistent with a
single value of photon index (1.345$\pm$0.010).  \cite{Hirayama et
al.(2002)} analyzed the {\sl ASCA} data of PSR B0540-69 and
suggested that the spectrum is harder at pulse peak.

In this paper, using the archive  {\sl RXTE} data, we present the
high quality X-ray pulse profiles in different energy ranges and the
phase-resolved spectra of PSRs B0531+21, B1509-58 and B0540-69. Even
with a lot of existing phase-resolved spectral analyses of the Crab
pulsar, detailed studies are still necessary, especially for the
spectral variation of inter-peak region and the turning points of
the two pulse peaks. Previously, the phase resolved X-ray
spectroscopy of PSRs B1509-58 and  B0540-69 was only done by
dividing their pulses into a few phase bins each. In this work, we
will use most of the archival {\sl RXTE} data of these three pulsars
to obtain their high precision X-ray light curves and the phase
resolved spectra, which will supply more information for
understanding the magnetosphere structure and set more constraints
for the theoretical models. We will also check whether there exist
spectral evolutions at a timescale of about ten years.

\section{Observation and Data Reduction}

The data we analysed in this paper were obtained by the Proportional
Counter Array (PCA) on board {\sl RXTE}. PCA is composed of five
Proportional Counter Units (PCUs), which has effective energy
coverage of 2 to 60\,keV, a total collection area of 6500\,cm$^2$, and
the best ever time resolution of up to 1$\mu$s (in Good Xenon
mode) \citep{Jahoda et al.(1996)}. The large detection area and the
superb time resolution make PCA an ideal instrument to study the
detailed temporal and spectral properties of pulsars.

{\sl RXTE} has made hundreds of observations on the three young
pulsars that we studied, which were listed in Table.
\ref{table:ListObsID}. The total exposure time of PSRs B0531+21,
B1509-58, and B0540-69 is 172, 578, and 2,663 ks, respectively
(Table. \ref{table:ListObsID}), and with such plenty of data we can
perform detailed phase-resolved spectroscopic study. The Event mode
data were selected for our analyses, so as to have good time
resolution and enough spectral channels.

Heasoft (v6.10) was used to process the observational data as
follows: (1) Create the filter file for each observation with
XTEFILT; (2) generate the Good Time Interval (GTI) file  by MAKETIME
using the filter file; (3) using the GTI file to create the ``good''
events file with GROSSTIMEFILT; and (4) remove the clock events from
the event mode data by SEFITER and FSELECT. The resulted new events
data will be used in the timing and spectral analyses.

First, we study the timing properties of the three pulsars with the
above observations. The arrival time for each photon was converted
to the Solar System Barycentre using ephemeris DE200 with FAXBARY.
For each observation we obtained the period of a pulsar by folding
the observed counts to reach the maximum Pearson $\chi^2$. If
the integration time for an observation is not long enough, a few
neighboring observations will be merged in the timing analysis.
The rotation frequencies of the three pulsars are fitted by a quadratic
polynomial:
\begin{equation}
\nu(t)=\nu_{0}+\dot{\nu}(t-t_{0})+\frac{1}{2}\ddot{\nu}(t-t_{0})^{2}.
\label{eq:1}
\end{equation}

In order to study the variation of the pulse profile versus photon
energy, we create the pulse profiles of the three pulsars in
different energy ranges as follows. (1) Fold the observed counts
for each observation of a pulsar within a certain energy range
into a light curve and subtract the baseline using the mean of
phase ranges 0.63-0.83, 0.8-1.0, and 0.8-1.0 for PSRs B0531+21,
B1509-58, and B0540-69, respectively; (2) Align and sum up the
baseline subtracted light curves in the same energy range of a
pulsar; (3) Normalize the light curves to make the peak value equal
to 1. The numbers of the light curves of these three pulsars are 211,
249, and 608, respectively. The final light curves are then used
for the studies of the evolution of the profiles versus energy.

We produce the phase resolved spectra, the response matrixes, and the
pulse profiles observation by observation and merge them together according to
their respective phase bin for the final profile and spectral analyses. The
phase resolved spectra were created by FASEBIN of {\sl ftools}, and the
baseline (unpulsed contribution) was subtracted with FBSUB. The
number of phase bin (the NPH parameter for the FASEBIN script, NPH: number of phase bin.) for
PSRs B0531+21, B1509-58, and B0540-69 are 200, 50, and 25,
respectively and some phase bins will be merged into one bin
to guarantee the quality of the spectrum. As there is precise radio
ephemerides (including periods and phases) for PSR B0531+21 at
ATNF\footnote{ http://www.atnf.csiro.au/resear
ch/pulsar/archive/data.html}, they were used in the process to
produce the phase resolved spectra of this pulsar. For PSRs B1509-58
and B0540-69, since the radio ephemerides are not available for the
whole observation duration, the ephemerides derived from the X-ray results
were used to gain the phase-resolved spectra, and phase 0 of this observation was
obtained by a cross-correlation analysis between its light curve and
the reference one. We chose the flat regions of the pulse profiles
as the unpulsed background phases (Fig. \ref{fig4},
Fig. \ref{fig12}, and Fig. \ref{fig13}). The unpulsed
background phase ranges that correspond to the PHASEMIN and
PHASEMAX parameters in FBSUB are 0.63-0.83, 0.8-1.0,
and 0.8-1.0 for the three pulsars, as mentioned above. The response matrix
for a spectrum was produced with PCARSP. The command to
merge the spectrum is ADDSPEC and that to merge the response
matrixes is ADDRMF. The weights used in ADDRMF are determined
by the total photon counts of the pulse in the corresponding
observations. Because PSR B0531+21 is bright, we obtained 42 merged
spectra for each phase bin, while for PSR B1509-58 and B0540-69,
because they are much fainter, only one merged spectrum has been
obtained for each phase bin. The pulse profiles in different energy
ranges were gained from the phase-resolved spectra created by
the FASEBIN and FBSUB of {\sl ftools} with more phase bins.

All the spectra were analysed with XSPEC {\bf (ver 12.6.0)}. The
spectral model is power law with photoelectric absorption, i.e.
[pha*(pl) in XSPEC]. The absorption column density ($N_{H}$) of PSRs
B0531+21, B1509-58, and B0540-69 were fixed at 0.36, 1.27,
0.46$\times10^{22}$ cm$^{-2}$ \citep{Massaro et al.(2000), Marsden
et al.(1997), Kaaret et al.(2001)}. For the Crab pulsar, the spectra
were also analysed with log-parabola model
\citep{Massaro et al.(2000),Massaro et al.(2006)}.
The systematic error was set as 0.01.
The errors of photon indices are at 90$\%$ confidence
level and the errors of the rest data are at
68.3$\%$ confidence level.

\section{Results}

\subsection{Crab}

\subsubsection{Timing and pulse profiles of the Crab pulsar}

\begin{figure*}
\begin{center}
\includegraphics[width=1\textwidth]{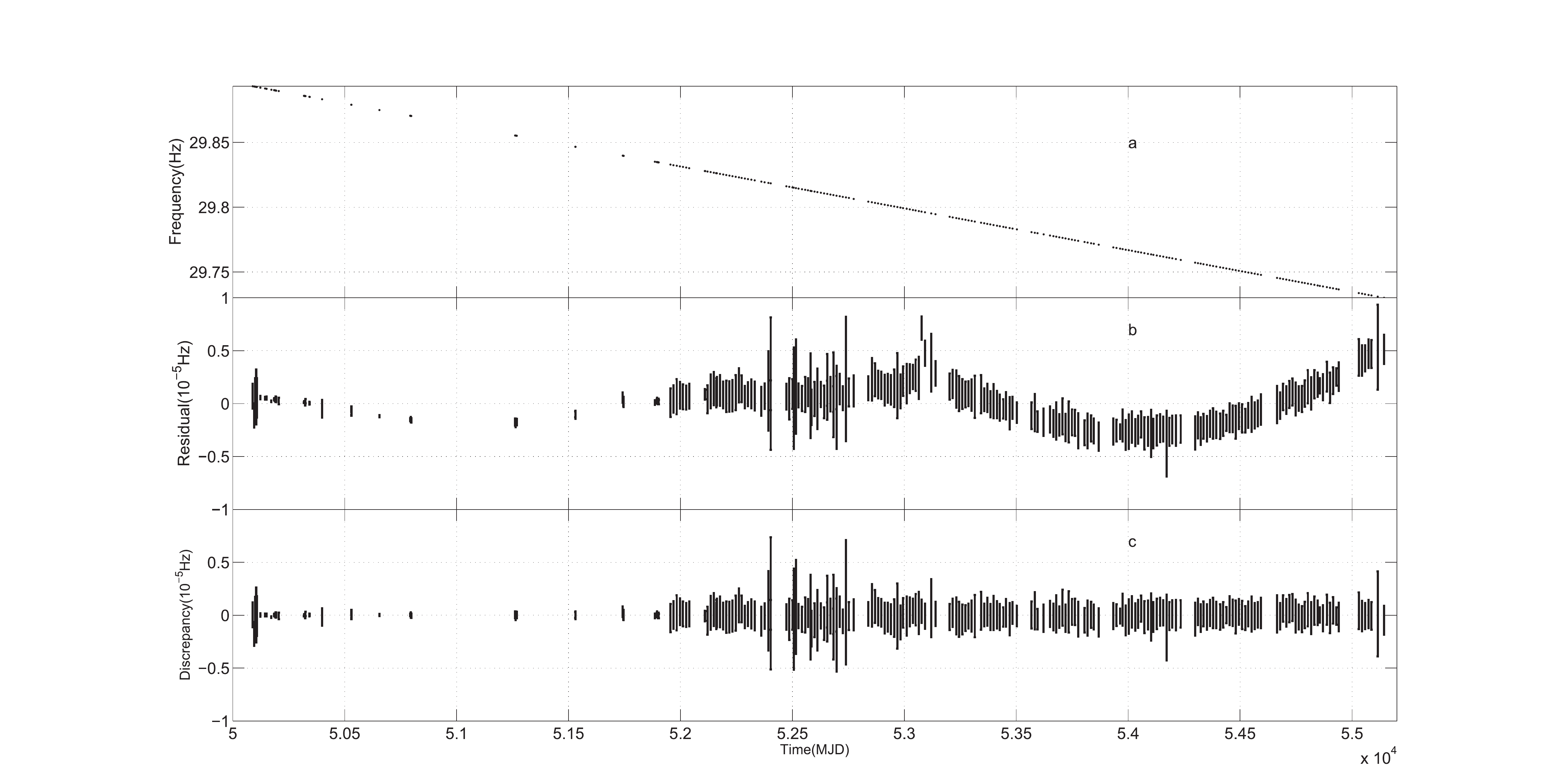}
\caption{The timing result of the Crab pulsar. Panel $a$ shows
the frequency evolution, panel $b$ shows the frequency residuals of
the fit with equation(1).  and panel $c$ shows the differences  between the
frequencies measured by X-ray observations and those calculated from the
radio ephemeris.
\label{fig1}}
%\end{center}
%\end{figure}
%
%\begin{figure}
%\begin{center}
\includegraphics[width=0.8\textwidth]{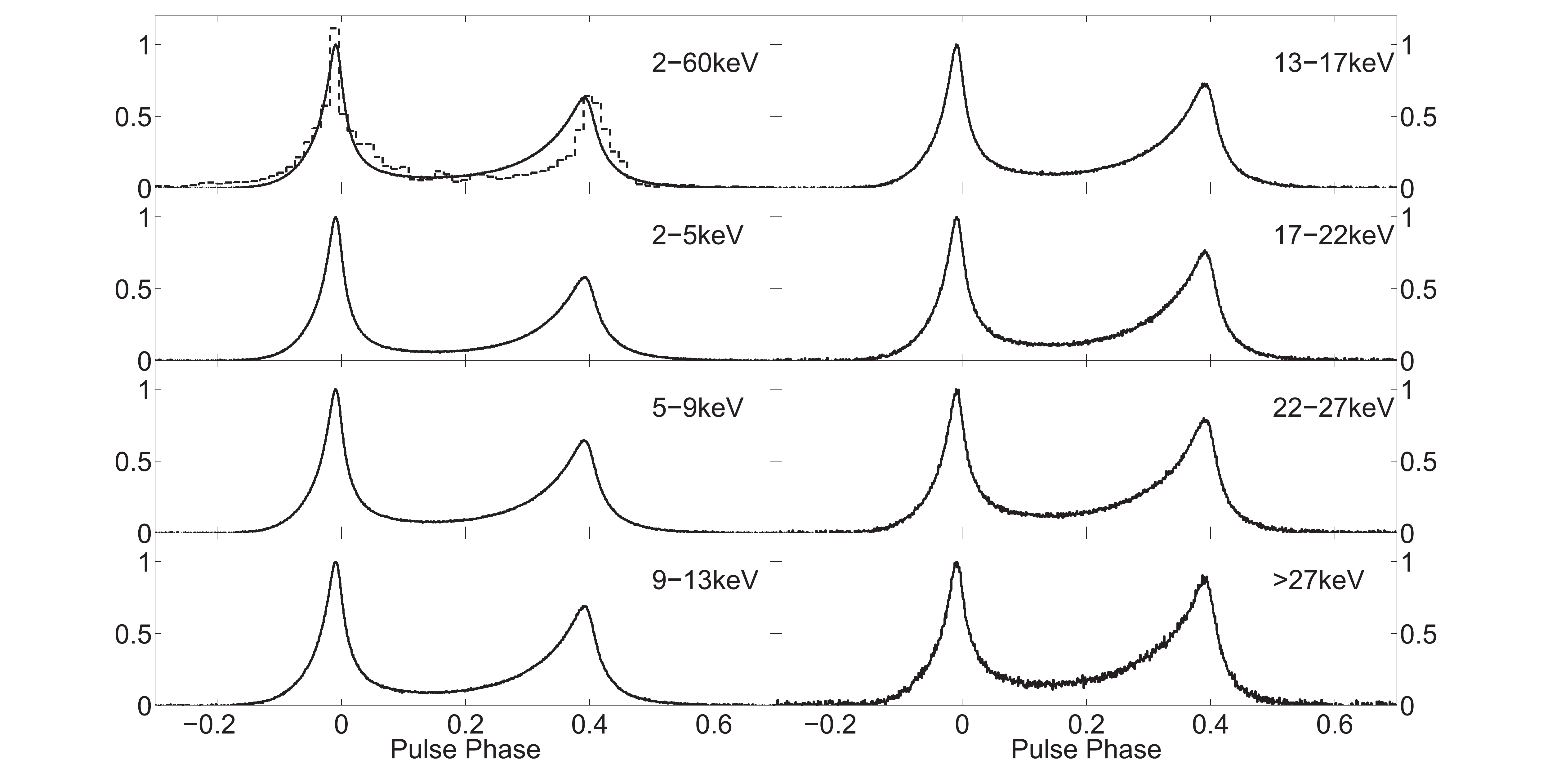}
\caption{The X-ray profiles of the Crab pulsar in different energy
ranges. Phase 0 represents the position of the main radio peak.
The dashed line represents the result of the theoretical model
\citep{Jia et al.(2007)}.\label{fig4}}
\end{center}
\end{figure*}

To study the timing behaviors of the Crab pulsar we fit its rotation
frequencies by a quadratic polynomial (equation (\ref{eq:1})).
Frequencies and residuals are plotted in Fig. \ref{fig1}a and Fig.
\ref{fig1}b respectively, and the fitting results are listed in Table 2.
The difference between the X-ray and
the radio frequencies are displayed in Fig. \ref{fig1}c
The later is from Jodrell Bank Observatory
\footnote{http://www.jb.man.ac.uk/research/pulsar/crab.html}. It is
clear that several glitches occurred in this duration, and the X-ray
derived frequencies are consistent with the radio ephemerides.
Therefore, the radio ephemerides are used when we produce the X-ray pulse
profiles in different energy bands and the phase resolved X-ray
spectra.

The profiles of the Crab pulsar are created by FASEBIN and FBSSUM of
ftools. For each photon event, the RXTE clock corrections and the
barycenter corrections based on the JPL DE-200 solar system ephemeris are applied.
Then the absolute phase is calculated by referencing to the radio timing ephemeris.
Therefore, for an X-ray pulse profile in a specific energy range created
by FBSSUM, the phase is the relative phase to the main radio peak.
Because a large number of photons have been
collected from the Crab pulsar, the period is divided into
1000 phase bins so as to show the detailed structure of the light
curve. The energy ranges used to create the light curves are
2-60, 2-5, 5-9, 9-13, 13-17, 17-22, 22-27, and $>$27\,keV,
corresponding to PCA channels 0-255, 0-10, 11-20, 21-30,
31-40, 41-50, 51-60, and 61-255, respectively. The resulted
light curves are plotted in Fig. \ref{fig4}, where phase 0
represents to the position of the main radio peak as mentioned
above. Obviously, the shape of the light curve varies with energy.

In order to obtain the phases of the two X-ray peaks more
accurately than the size of the phase bin,
we fit them with an empirical formula proposed by
\cite{Nelson et al.(1970)}.
\begin{equation}
L(\phi-\phi_{0})=N\frac{1+a(\phi-\phi_{0})+b(\phi-\phi_{0})^{2}}{1+c(\phi-\phi_{0})+d(\phi-\phi_{0})^{2}}e^{-f*(\phi-\phi_{0})^{2}}+l,
\label{eq:5}
\end{equation}
where $L$ is the intensity at phase $\phi$, $l$
the baseline of the light curve, $\phi_0$ the phase shift,
$N$ the pulse height of the profile, and $a$, $b$, $c$, $d$ and $f$
the shape coefficients. The pulse phase is measured in phase units,
range (0,1). The coefficients of the two peaks were
fitted separately, since the width of the main peak broadens
with energy increasing \citep{Mineo et al.(1997),Massaro et al.(2006)},
which is verified in this work and the shapes of the second peak
are different from the main peak.

\begin{figure*}
\begin{center}
\includegraphics[width=0.8\textwidth]{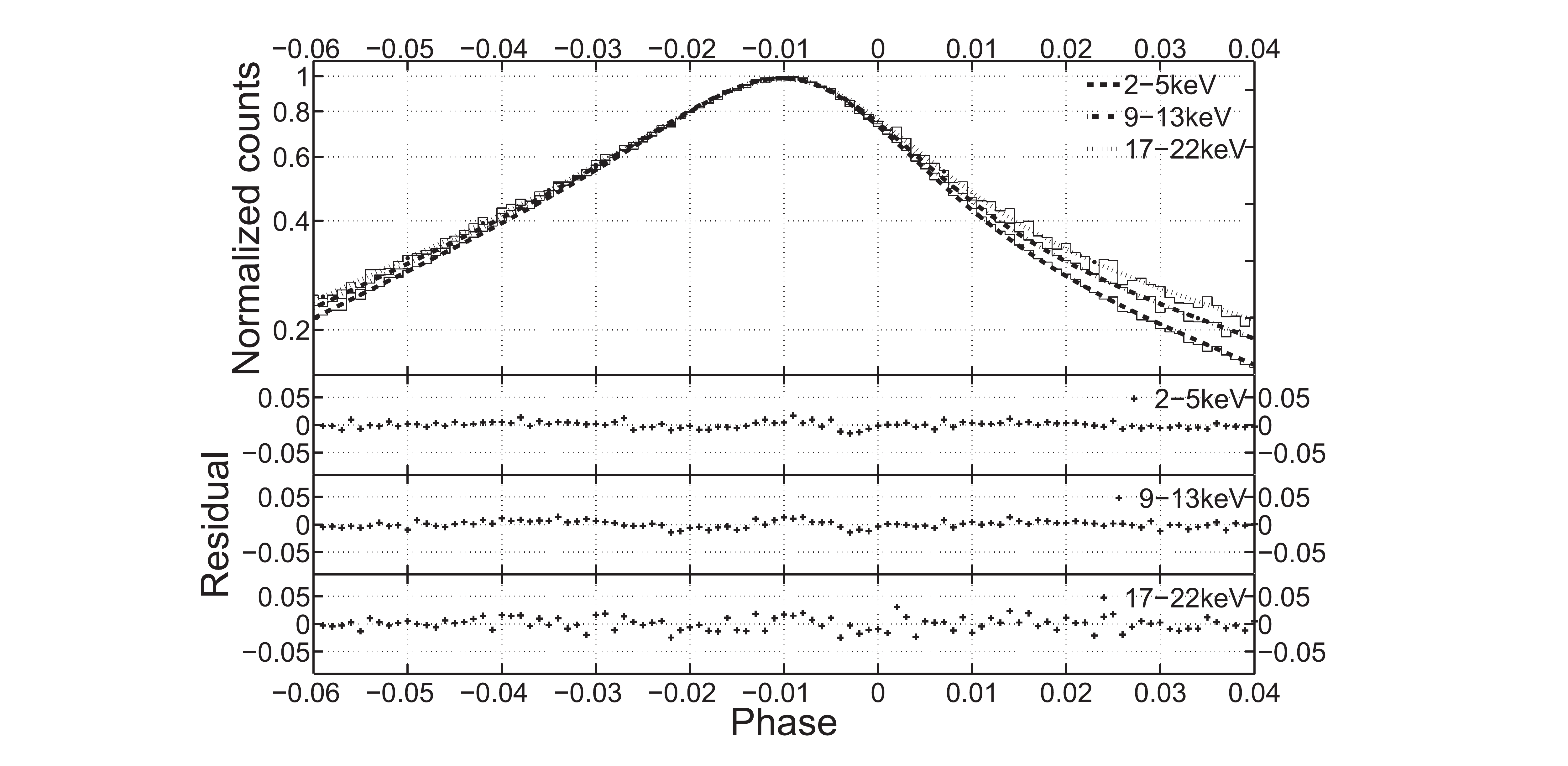}
\caption{
The shape of the main peak of the Crab profile in
three energy ranges: 2-5\,keV, 9-13\,keV,
17-22\,keV. Phase 0 represents the position of the main radio peak.
The dashed line, dot-dashed line, and dotted line
represent the fitted profiles with Nelson's formula separately
in these three energy ranges.
\label{fig41}}

\includegraphics[width=0.8\textwidth]{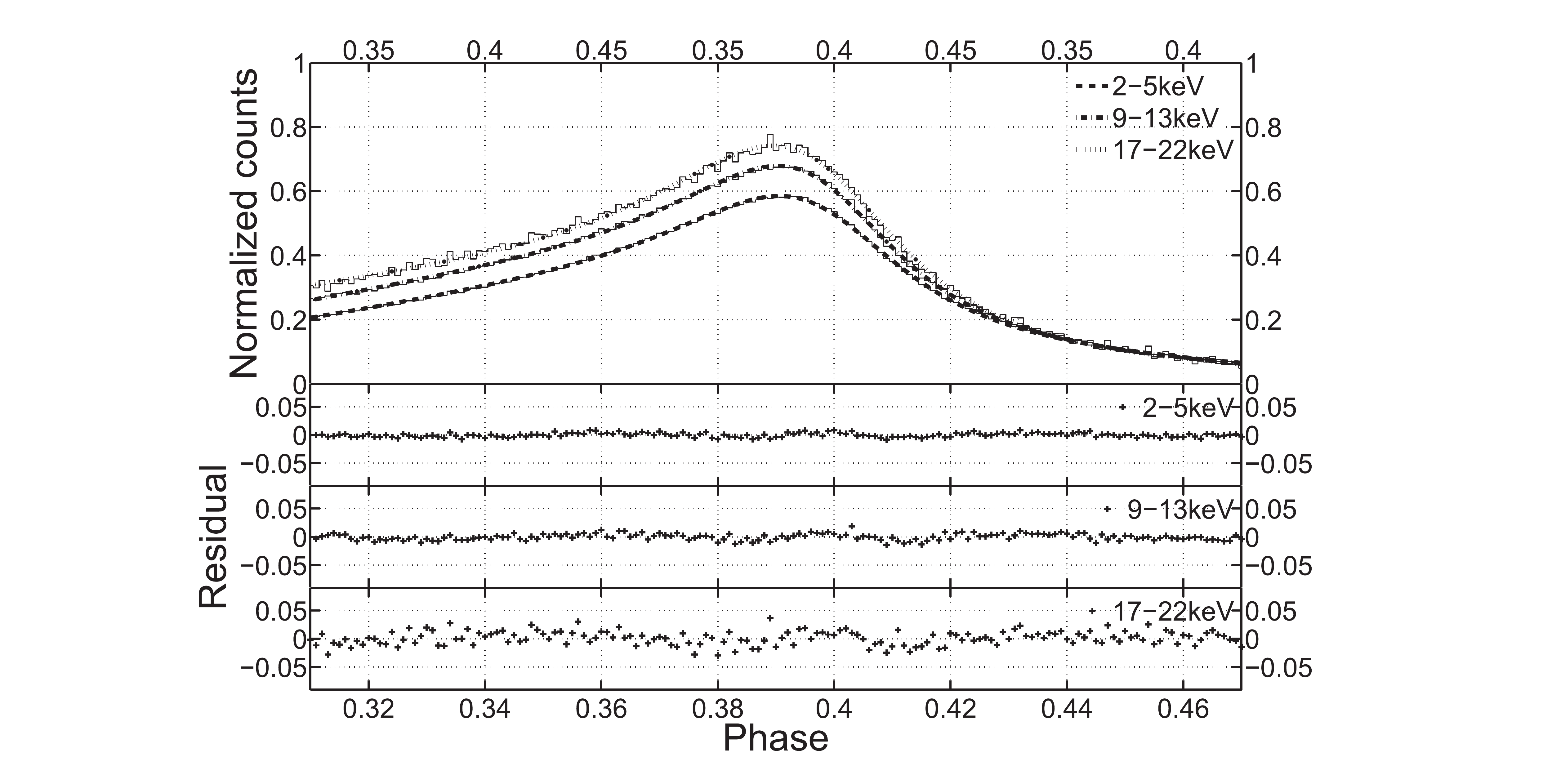}
\caption{The shape of the second peak of the Crab profile in
three energy ranges: 2-5\,keV, 9-13\,keV,17-22\,keV.
Phase 0 represents the position of the main radio peak.
The dashed line, dot-dashed line, and dotted line
represent the fitted profiles with Nelson's formula separately
in these three energy ranges.
\label{fig42}}
\end{center}
\end{figure*}

\begin{figure*}
\begin{center}
\includegraphics[width=0.8\textwidth]{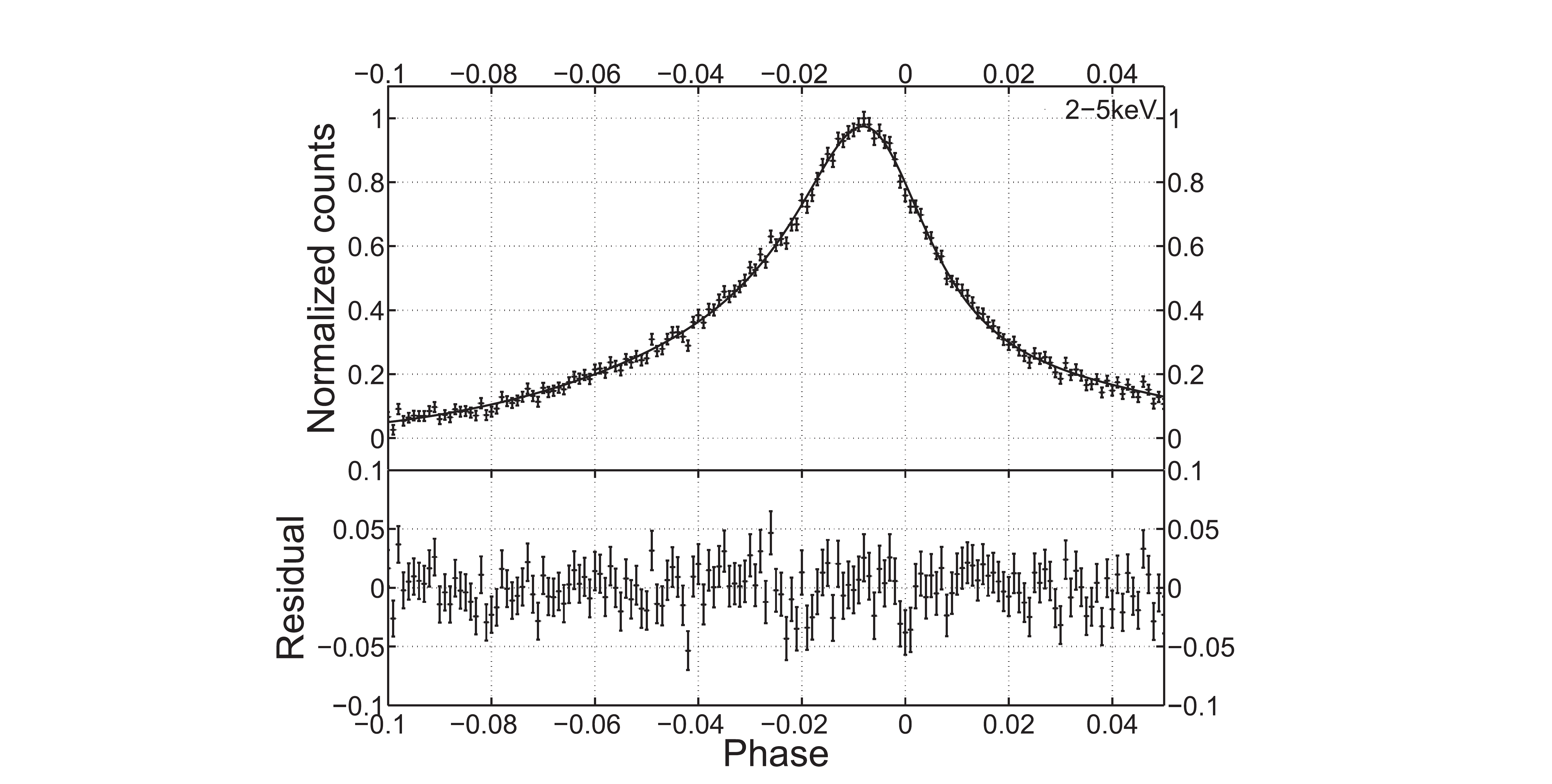}
\caption{ Upper pannel: The fitted profile of the main peak of
the Crab profile with fixed parameters of Nelson's formula
for observational ID 60080-01-03-00(profile phase range:-0.29 to 0.07).
Lower pannel: The residuals between the observational profile
and fitted profile. The reduced $\chi^{2}$ is 0.99 with d.o.f. 358.
\label{fig43}}

\includegraphics[width=0.8\textwidth]{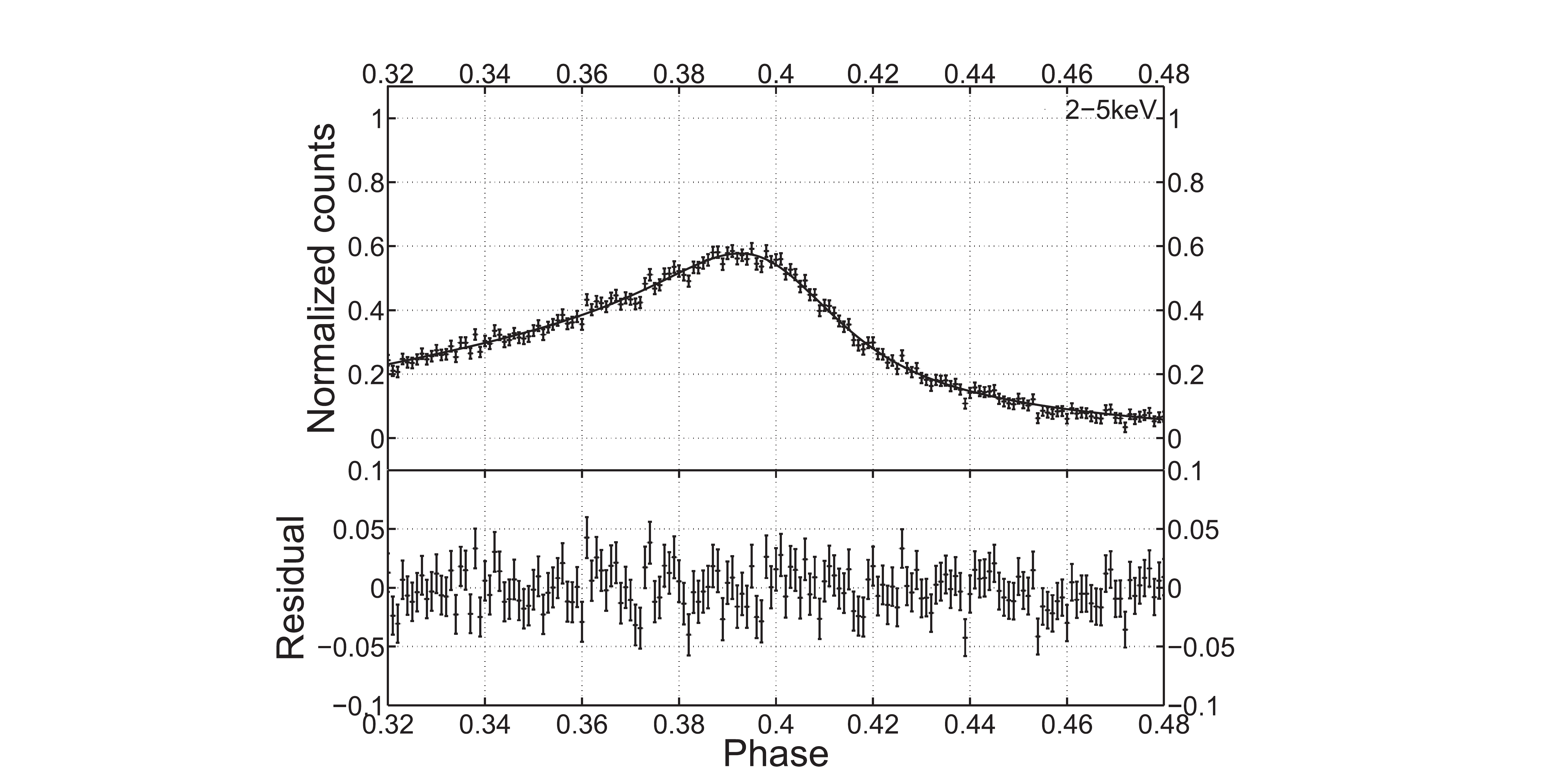}
\caption{Upper pannel: The fitted profile of the second
peak of the Crab profile with fixed parameters of Nelson's formula
for observational ID 60080-01-03-00(profile phase range: 0.32 to 0.71).
Lower pannel: The residuals between the observational profile
and fitted profile. The reduced $\chi^{2}$ is 1.00 with d.o.f. 388.
\label{fig44}
}
\end{center}
\end{figure*}

\begin{figure*}
\begin{center}

\includegraphics[width=0.8\textwidth]{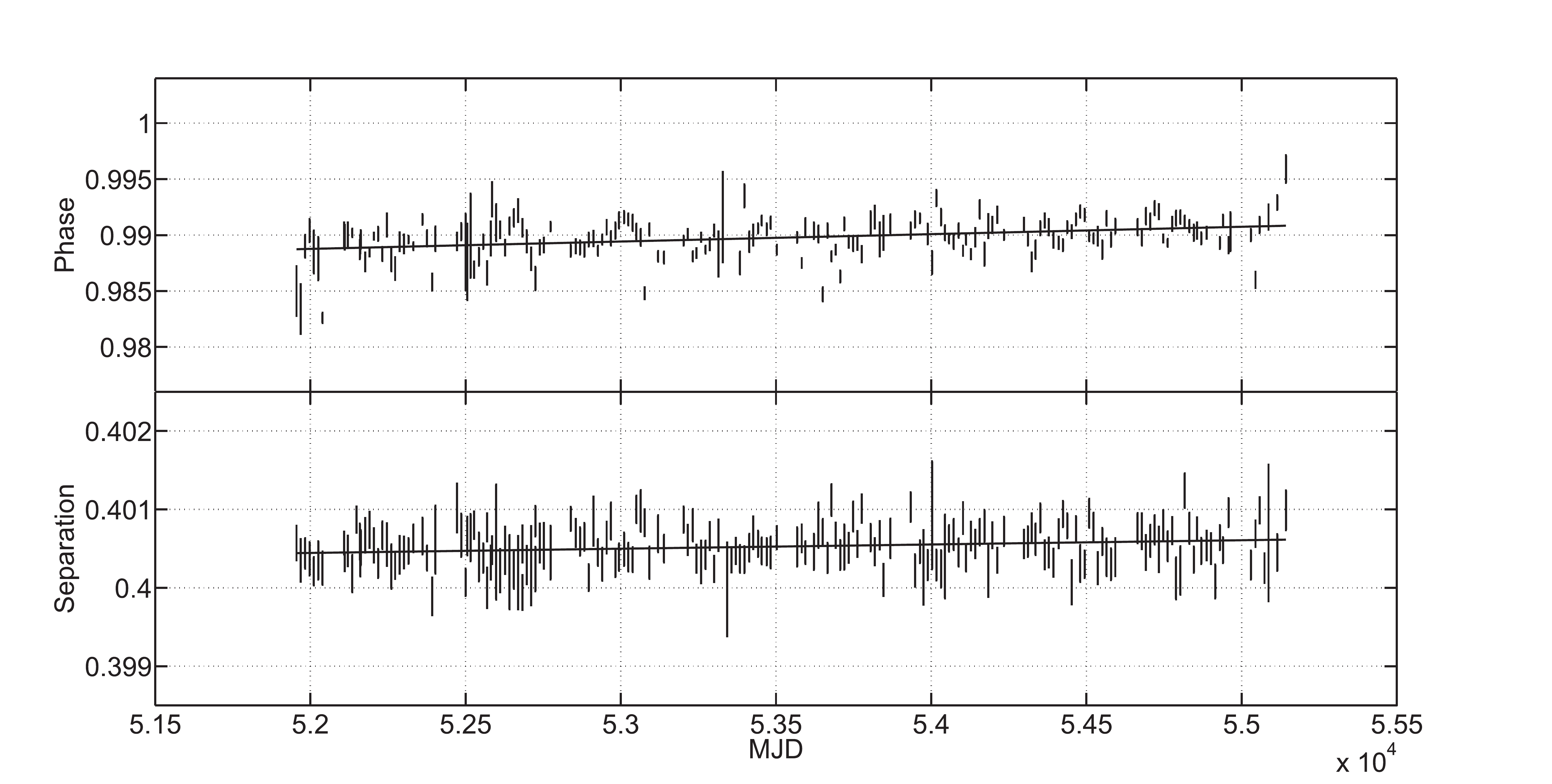}
\caption{
Upper pannel: The phase of the X-ray main peak of the Crab pulsar
relative to the radio phase from the Jordell Bank radio ephemeris,
in while the phase of the radio main peak is 1.0.
Lower pannel: The separation of the two X-ray peaks in different epoches.
\label{fig5}}

\includegraphics[width=0.8\textwidth]{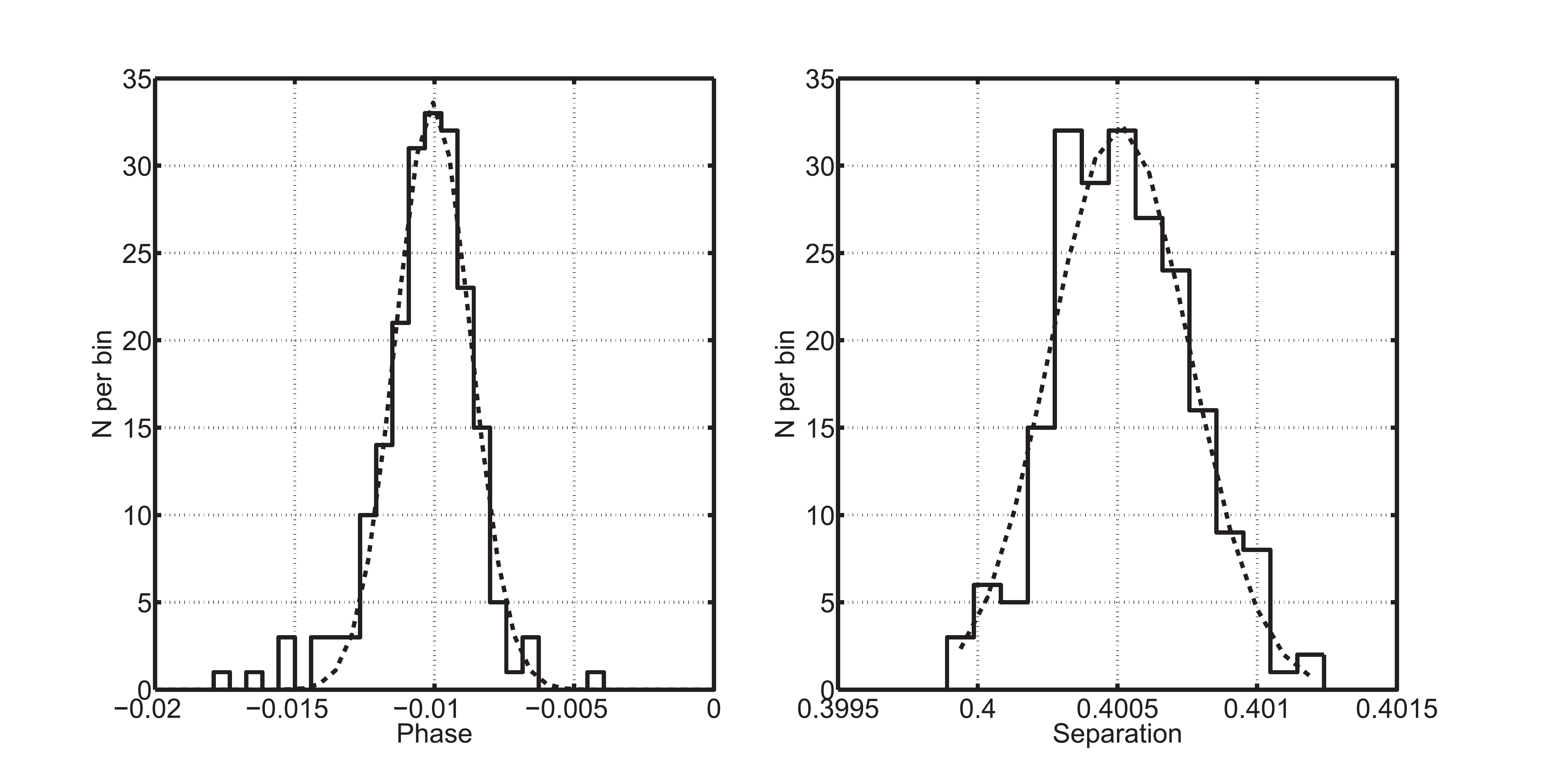}
\caption{
Left pannel: The distribution of the phase-lag between the X-ray main
peak and the radio main peak of the Crab pulsar with bin size
$5.8\times10^{-4}$ periods. The dashed line represents
the fitted Gaussian function with $\mu=-0.0101$
and $\sigma=1.3\times10^{-3}$. Right pannel: The
distribution of the separation between the two X-ray peaks with
bin size $9.6\times10^{-5}$ periods. The dashed
line represents the fitted Gaussian function with $\mu=0.4005$
and $\sigma=2.5\times10^{-4}$.
\label{fig45}}

\end{center}
\end{figure*}

The fitting procedures of the two peak profiles can be divided into
two steps: (1) Obtain the coefficients $a$, $b$, $c$, $d$, $f$ , as
well as  $N$, $l$, and $\phi_{0}$ from the fitting with the Nelson's
formula to the profiles (Fig. \ref{fig4}) accumulated from all the
observations. The fitted profiles of the main peak and the second
peak in three energy ranges: 2-5\,keV, 9-13\,keV, and 17-22\,keV are
displayed in Fig. \ref{fig41} and Fig. \ref{fig42}, respectively.
The final shape parameters of the two peaks in different energy
ranges are listed in Table. \ref{table:nelsonpar}; (2) Each of the
211 observed profiles in a specific energy range is fitted with the
Nelson's formula coefficients of which are fixed except $N$,
$\phi_{0}$, and $l$. The comparisons of one observational profile with
the fitted profiles for the two peaks are shown in Fig. \ref{fig43}
and Fig. \ref{fig44}. The reduced $\chi^{2}$ of the main peak and
the second peak are 0.99 (d.o.f. 358) and 1.00 (d.o.f. 388), showing
that the profiles could be properly fitted. The position of the
maximum of the fitted profile is taken as the phase of the main
peak, while the separation of the positions of the two maxima is
taken as the separation of the two peaks. Fig. \ref{fig5} plots the
211 phase-lags and separations, in which the errors of the
phase-lags are a combination of the main peak phase errors and the
rms deviations of the radio timing
ephemerides\footnote{http://www.jb.man.ac.uk/research/pulsar/crab.html},
while the errors of the separations are obtained from the errors of
the phases of the two peaks. The phase errors of the X-ray main and
second peaks used in the above calculation are obtained from the
parameter $\phi_{0}$ in the fitting processes.

Averaging the fitting results to the 211 light curves in 2-60 keV,
we found that the phase of the X-ray main peak (2-60\,keV) leads the
radio main peak with a mean value of 0.0101$\pm$0.0001 periods (Fig.
\ref{fig5}) and the mean separation between the main peak and the
second peak is 0.40051$\pm$0.00003 periods. Here the error of
the phase-lag between the X-ray main peak and the radio main peak is
estimated from the distribution of the phase-lag values of the 211
X-ray light curves, and the error of the separation between two
X-ray peaks is obtained similarly. As shown in the left panel of
Fig. \ref{fig45}, the 211 phase-lags distribute as a Gaussian shape.
We then fit the distribution with a Gaussian function, with three free
parameters: namely the normalization G, the mean $\mu$, and the standard
deviation $\sigma$. $\sigma$ is regarded as the standard deviation of the
phase-lag measured from one light-curve, and the standard error of
the mean phase-lag is estimated by $\sigma/\sqrt{N}$, where $N=211$,
the number of the phase-lags from which the mean is obtained. The
fitted mean phase-lag is $\mu=-0.0101$ and
$\sigma=1.3\times10^{-3}$, while the mean phase of the X-ray main
peak(2-60\,keV) leads the radio main peak by 0.0101$\pm$0.0001
periods. Similarly, as illustrated in the right panel of Fig.
\ref{fig45}, the mean separation of the two X-ray peaks is
0.40051$\pm$0.00002 periods. We find that the separations between
the two peaks are more concentrated comparing with the phase-lag
distribution, implying that the errors of the phase-lags are
dominated by the rms deviations of the radio data.

\begin{figure*}
\begin{center}
\includegraphics[width=0.8\textwidth]{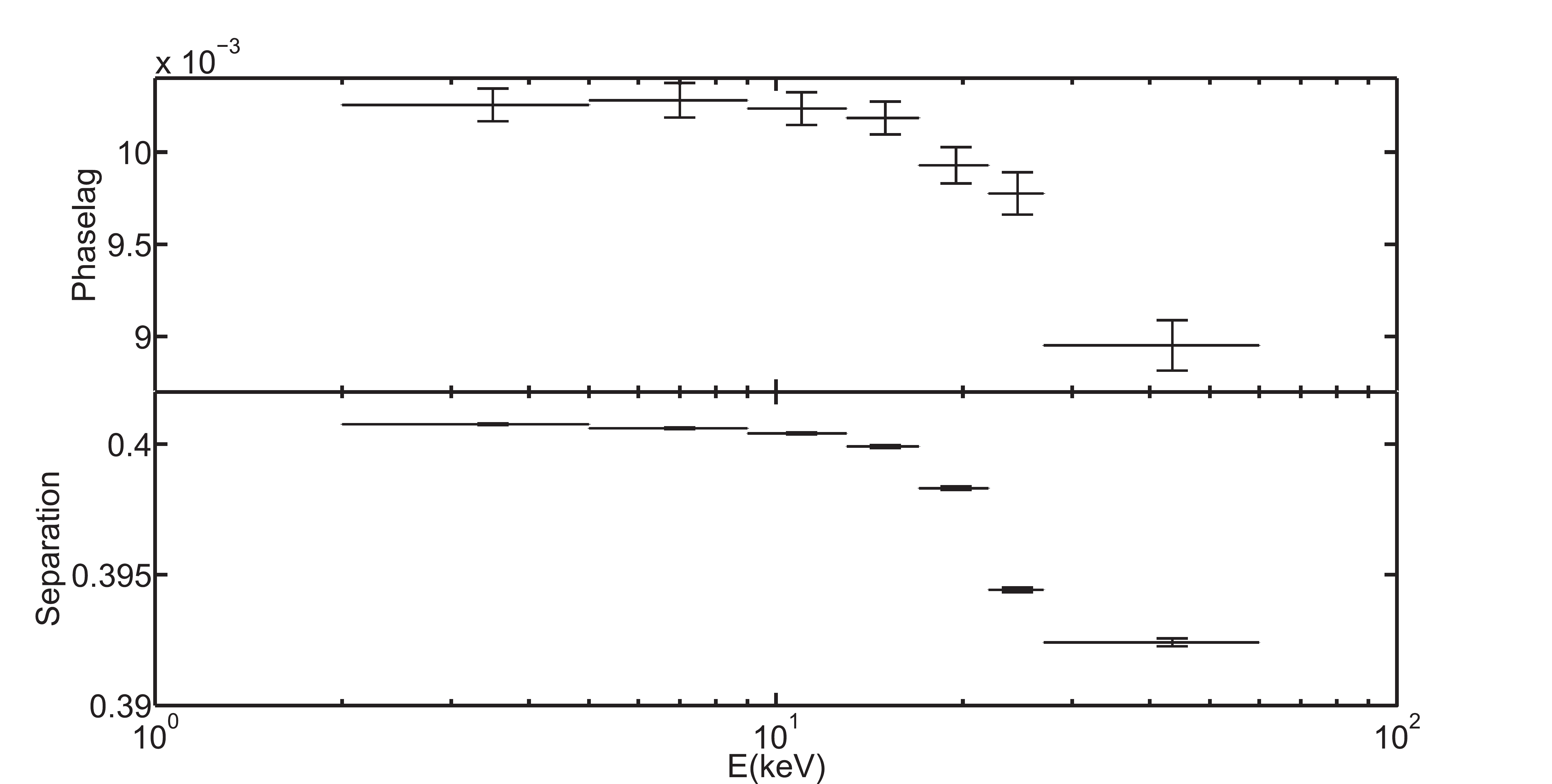}
\caption{
Upper pannel: The phase difference between the X-ray main
peak of the Crab pulsar to the radio main peak as a function of
X-ray photon energy. Lower Panel: The separation of the two X-ray
peaks as a function of energy.
\label{fig6}}

\includegraphics[width=0.5\textwidth,angle=0]{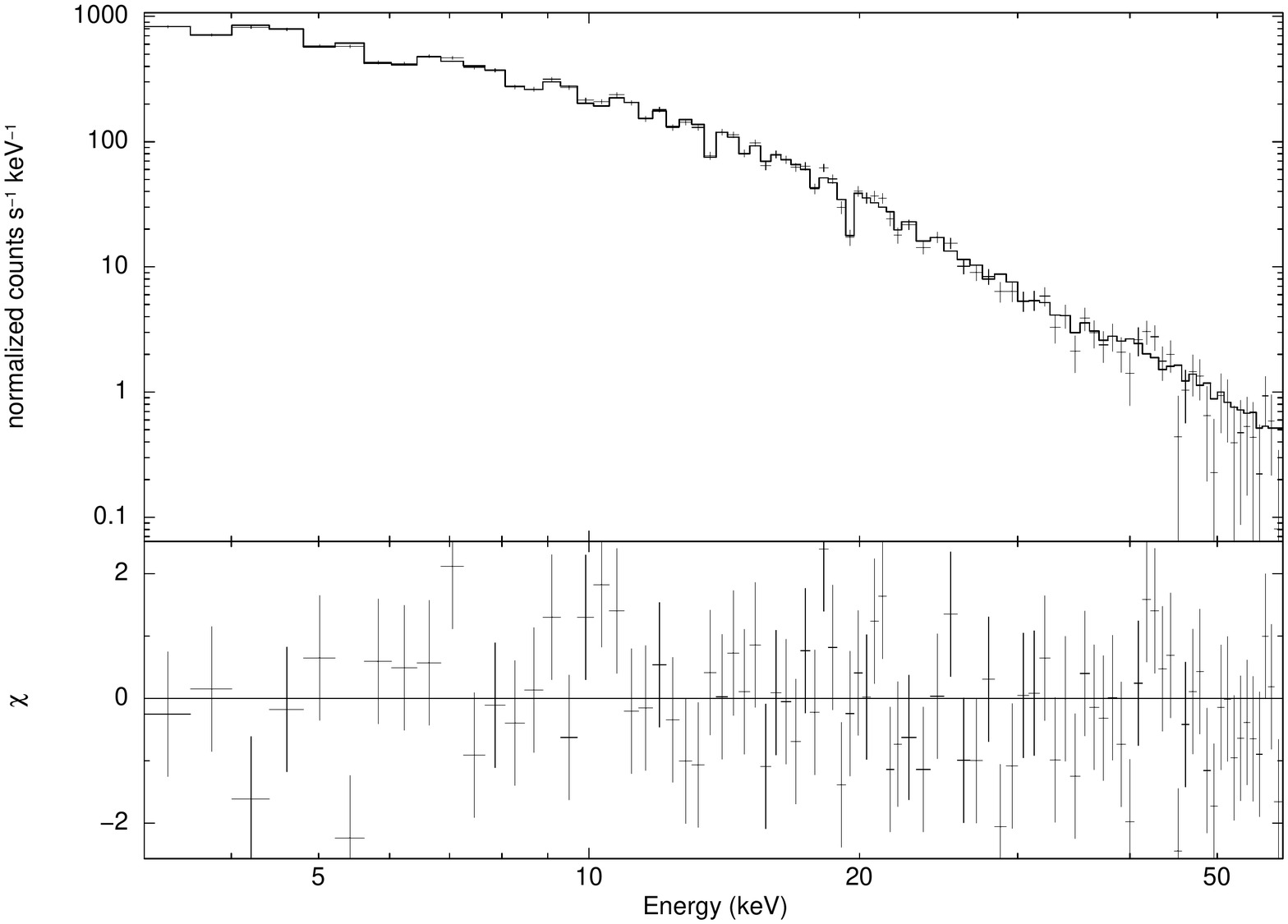}
\caption{The spectral fit of the Crab pulsar with a absorbed power law
($N_{H}=0.36\times10^{22}cm^{-2}$) at phase:0.0-0.005.
Upper pannel: The crosses represent the observations, and the solid line
is the model convolved with the response matrices. Lower pannel:
The residuals between the observation data and the best fitted model.
The reduced $\chi^{2}$ is 0.99 with the degree of freedom 86.
\label{fig111}}
\end{center}
\end{figure*}

Previous studies suggested that the phase-lag of the X-ray main peak
to the radio main peak and the separation of the two X-ray peaks
vary with energy and might change slightly with time \citep{Masnou
et al.(1994), Rots et al.(2004), Molkov et al.(2010)}. We found that
the phase lag of the radio peak to the X-ray
peak(2-60\,keV) shows an increasing trend with time, and a linear
fit gives a slope of (6.6$\pm$1.3)$\times$10$^{-7}$ periods
day$^{-1}$. The best fit slope is about two times larger than that
obtained by Rots et al. (2004), which is
(3.3$\pm$2.0)$\times$10$^{-7}$ periods day$^{-1}$. However, taking
the uncertainty into account, the two measurements are consistent
with each other. The separation between the two X-ray peaks does not
exhibit such a strong evolutionary trend with time. A linear fit to
the data gives a slop of (5.4$\pm$1.8)$\times$10$^{-8}$ periods
day$^{-1}$. The energy dependence of the phase-lag between the radio
and X-ray main peaks found by \cite{Molkov et al.(2010)} is
confirmed here (Fig. \ref{fig6}), which reaches the maximum at
5-9\,keV and then decreases rapidly. Such an evolutionary trend is
consistent with the multi-wavelength behaviors obtained by
\cite{Kuiper et al.(2003)} and \cite{Molkov et al.(2010)}. Fig.
\ref{fig6} also shows that the separation of the two X-ray peaks
decreases with the energy, which is consistent with the result of
\cite{Eikenberry and Fazio(1997)}. The errors of the phase-lags and
separations in different energy ranges are estimated similarly to
that in the last paragraph. Affected by the rms deviations of the
radio data, the errors of the phase-lags are almost the same with
each other as shown in the upper pannel of Fig. \ref{fig6}.

\subsubsection{Phase resolved spectral analyses}

\begin{figure*}
\begin{center}
\includegraphics[width=0.5\textwidth,angle=0]{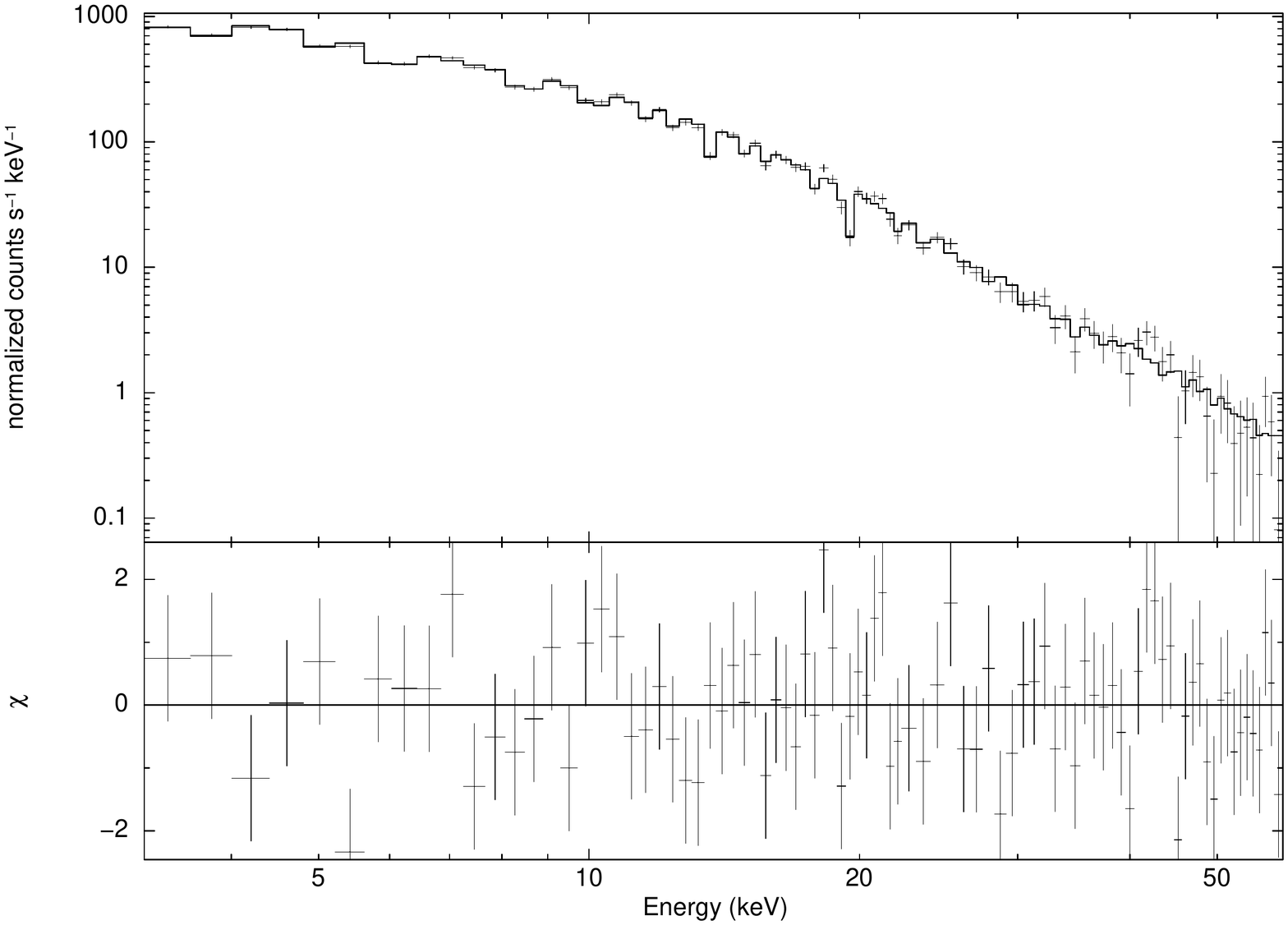}
\caption{Similar to Fig. \ref{fig111}, but the fitted model is log-parabola with $E_{0}$ fixed
at 1\,keV. The reduced $\chi^{2}$ is 0.93 with the degree of freedom 85.
\label{fig112}}

\includegraphics[width=0.8\textwidth]{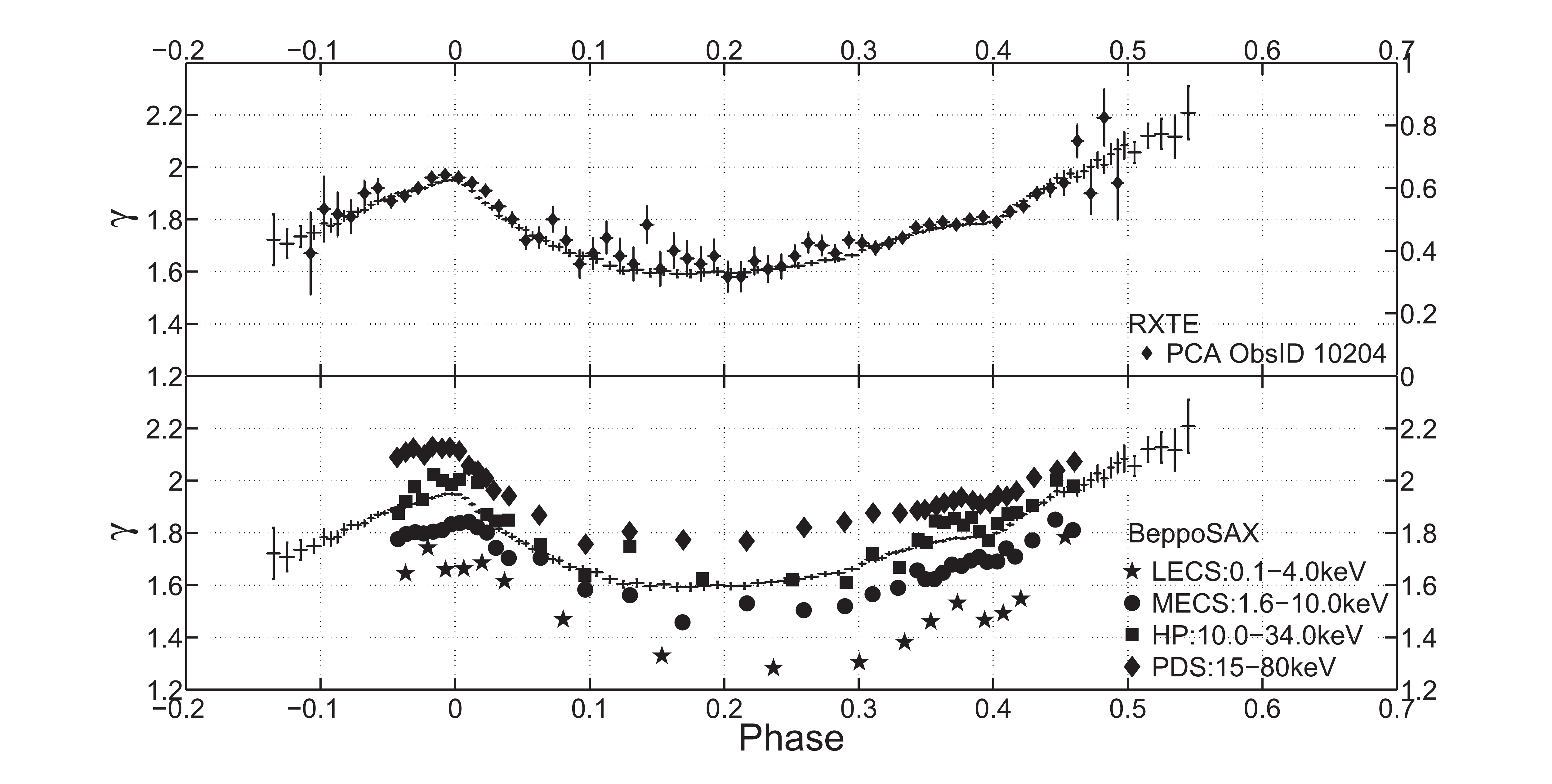}
\caption{Upper pannel: Comparison of the photon indices obtained in this work(``+'') and that
by Pravdo et al. (1997). Lower pannel: Comparison of the photon indices
obtained in this work and the results observed by the four instruments of {\sl BeppoSAX}.
\citep{Massaro et al.(2000)}
\label{fig114}}
\end{center}
\end{figure*}

Since plenty of data were used, we divided the X-ray pulse of the
Crab pulsar into 107 sections whose width is determined by the total
counts detected in this section, and 42 spectra were created for
each section from different observation time intervals. The mean number
of counts per bin is 1109.2 and the minimum number of counts per bin is
76.5 after background substraction with energy bins 88.
The 42 spectra were fitted with a power-law model and log-parabola
\citep{Massaro et al.(2000), Massaro et al.(2006)}
with photoelectric absorption, where the value of N$_{H}$ was fixed at
0.36$\times$10$^{22}$cm$^{-2}$ \citep{Massaro et al.(2006)}.
The free parameters of the power law model are the photon index ($\gamma$) and the
normalization ($C$):
\begin{equation}
F(E)=CE^{-\gamma}.
\label{eq:12}
\end{equation}
While the photon index $\gamma(E)$ in the log-parabola model varies with
energy and the normalization is $K$:
\begin{equation}
F(E)=K(E/E_{0})^{-(\alpha+\beta\,Log(E/E_{0}))},
\label{eq:12}
\end{equation}
\begin{equation}
\gamma(E)=\alpha+2\beta\,Log(E/E_{0}),
\label{eq:14}
\end{equation}
where $E_{0}=1$\,keV. Both models were used to fit the
phase-resolved spectra of the Crab pulsar simultaneously
in order to compare the differences between them.
The comparisons between the two models fitting
the spectrum at phase: 0.0-0.005 are shown in Fig. \ref{fig111}
and Fig. \ref{fig112}. The result shows that
power-law model and log-parabola model were both suitable
to fit the spectra of the Crab pulsar in the energy range of
3-60\,keV. The mean of the 42 $\gamma$ values weighted by
photon numbers of the pulse, just as used in ADDRMF, was taken as
the final results of the photon index:
\begin{equation}
\gamma_{mean}=\sum w_{i}\gamma_{i}.
\label{eq:6}
\end{equation}
\begin{equation}
w_{i}=c_{i}/\sum c_{i},
\label{eq:7}
\end{equation}
where $c_{i}$ is the photon number of the pulse and $w_{i}$ is the weight.
The error of the mean photon index was obtained from the weighted sample variance.
It is the same for $C$ and parameters for log-parabola model. Data
points below 3\,keV and above 60\,keV were ignored for the spectral
analyses due to the low signal to noise ratios.

The fitted spectral parameters of the 107 pulse sections with the power
law model are listed in Table. \ref{table:Crabphrespec}, and the
photon indices are plotted in Fig. \ref{fig114}. It is clear that
the spectrum softens from the leading edge of the first peak to
the first intensity maximum, hardens through the inter peak and
softens again through the second peak. Such an evolutionary trend
is consistent with that reported by \cite{Pravdo et al.(1997)} and
\cite{Massaro et al.(2000)} (Fig. \ref{fig114}).
The photon indices of this study are close to the results of {\sl HP}
and between  the results of {\sl MECS} and {\sl PDS} of {\sl BeppoSAX}.
However, according to the higher accuracy of the present study, more
details are revealed: (1) The photon index changes smoothly, without any
small bump between the two peaks; (2) The hardest spectrum
($\gamma\sim1.60$) occurs in phase range 0.12-0.22, in which the
intensity also reaches the minimum between the two peaks; (3) The
spectrum softens slowly in the rising wing of the second peak and
the softening suddenly becomes much rapid from the maximum of the
peak. In order to describe the changes of the photon indices,
the pulse phase was divided into ten intervals and the relations of
photon indices with phase were fitted with a linear function in each
interval. The results are listed in Table. \ref{table:Crabphaseslope}.
The absolute slope values are different on
both sides of the main peak, which means the change is asymmetric.
For the second peak, the slope value of the tailing side is obviously larger
than that of the leading edge just as the previous description.
To integrate all the 107 spectra, the total absorbed flux of
Crab's pulse is (1.996$\pm$0.001)$\times$10$^{-9}$\,ergs s$^{-1}$
cm$^{-2}$ in the 2-10\,keV band.

\begin{figure*}
\begin{center}

\includegraphics[width=0.8\textwidth]{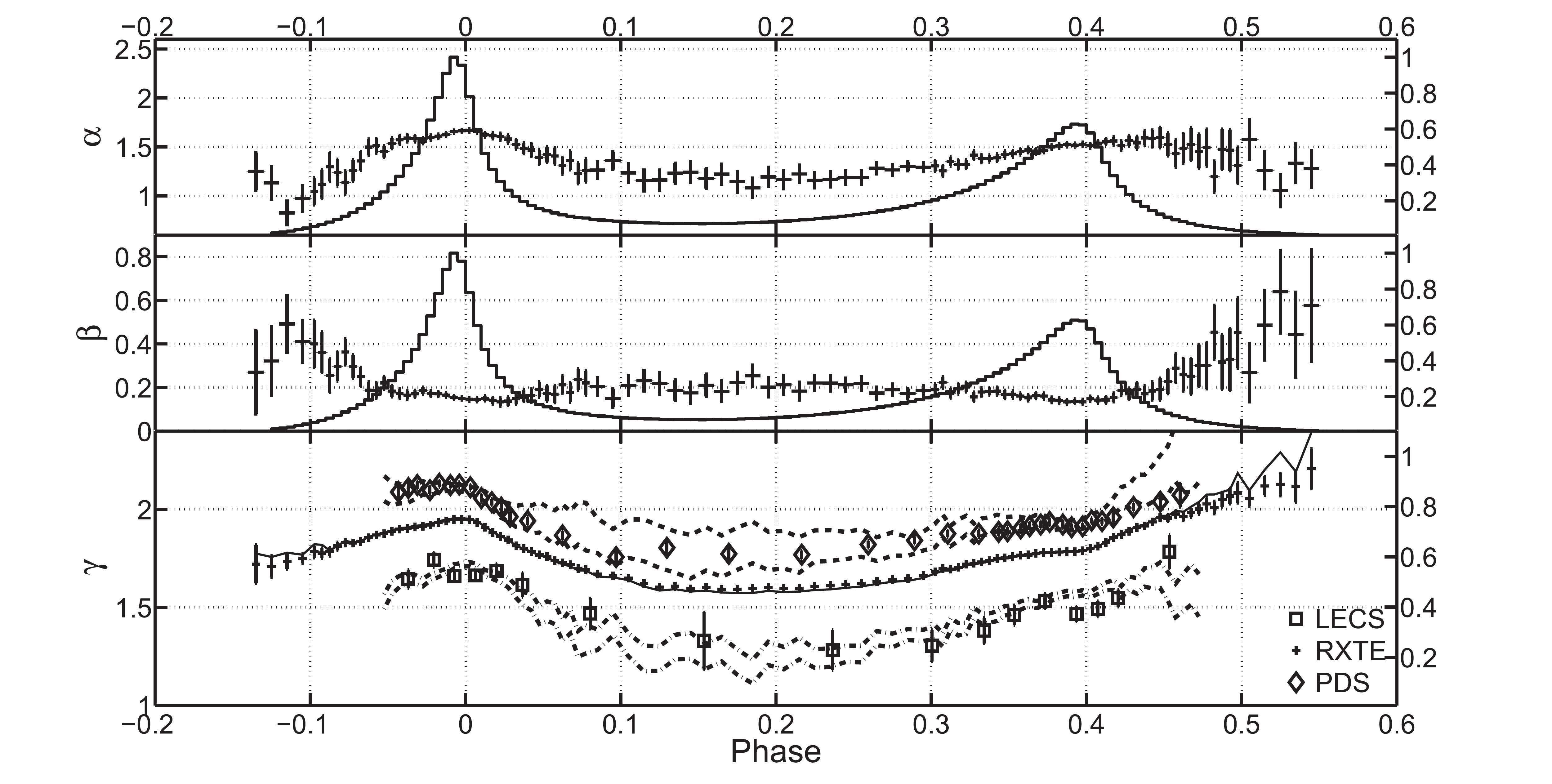}
\caption{
The results of the phase-resolved spectra of the Crab pulsar
with log-parabola model in different phase ranges.
Upper pannel: The parameter $\alpha$ in the log-parabola model.
Middle pannel: The parameter $\beta$ in the log-parabola model.
Bottom pannel: The comparison of the photon indices of the power law model and
spectral indices calculated from equation \ref{eq:14} where $E=1.5, 9.3, 30.0\,keV$.
The solid line represents the energy dependent photon indices for {\sl RXTE}.
The two dot-dashed lines represent $1\sigma$-uncertainties from
the extrapolation of the spectral distribution for LECS of {\sl BeppoSAX}.
The dashed lines are similar to dot-dashed lines but for
PDS of {\sl BeppoSAX}. The square, ``+'', and diamond points
represent the results of LECS, PCA, and PDS respectively.
\label{fig115}}

\end{center}
\end{figure*}

The results of the phase-resolved spectra with log-parabola
model are listed in Table. \ref{table:Crabphrespeclogpar}, and parameters
$\alpha$ and $\beta$ of the model in different phase ranges are plotted in
Fig. \ref{fig115}. The evolution of parameter $\alpha$ is similar to the
evolution of the photon index of the power law model except in
the trailing wing of the second peak. In the trailing wing of
the second peak, $\alpha$ becomes smaller with phase gradually
while the photon indices of the power law increase suddenly.
$\beta$ is lower at the two peaks, $\thicksim$0.16, moderate
at the interpeak region, $\thicksim$0.2, and higher in the leading
edge of the first peak(phase -0.14- -0.05) and the trailing edge of the
second peak(phase 0.45-0.55), $\thicksim$0.4. The positive value of the $\beta$
means that the photon indices in all phase ranges increase
with energy and larger $\beta$ means that the spectra softens
faster than in the other phase ranges. The result is consistent with the
conclusion of \cite{Massaro et al.(2000)} and \cite{Massaro et al.(2006)}
(Fig. \ref{fig114}). The energy dependent photon indices
can be calculated from equation \ref{eq:14} and its evolution
is similar to the evolution of the photon indices of the power
law model as shown in Fig. \ref{fig115} at $E=9.3$\,keV.

Using Ftest, we find that the log-parabola model is
more suitable than the power law model as to describing
the X-ray spectrum in each phase bin. As shown in
Table. \ref{table:Crabphrespeclogpar}, the substitution of
the power law model by the log-parabolic model is necessary at
a significance level of $>99.9\%$ in the two peak regions
(phase -0.03- 0.01 and phase 0.365-0.405). We also check whether
the extrapolation of the parabolic spectral model is consistent
with the observations in the adjacent energy ranges,
such as the results of LECS and PDS of {\sl BeppoSAX}.
As shown in Fig. \ref{fig115}, the results
of the extrapolation are consistent with the spectral
evolution obtained from LECS and PDS of {\sl BeppoSAX}
\citep{Massaro et al.(2000)}. To summarize, the log-parabola
model is suitable and better than the power law model for
the description of the X-ray spectra of the Crab pulsar.

\subsection{PSR B1509-58}

\subsubsection{Timing and Pulse profiles}

\begin{figure*}
\begin{center}

\includegraphics[width=0.8\textwidth,clip]{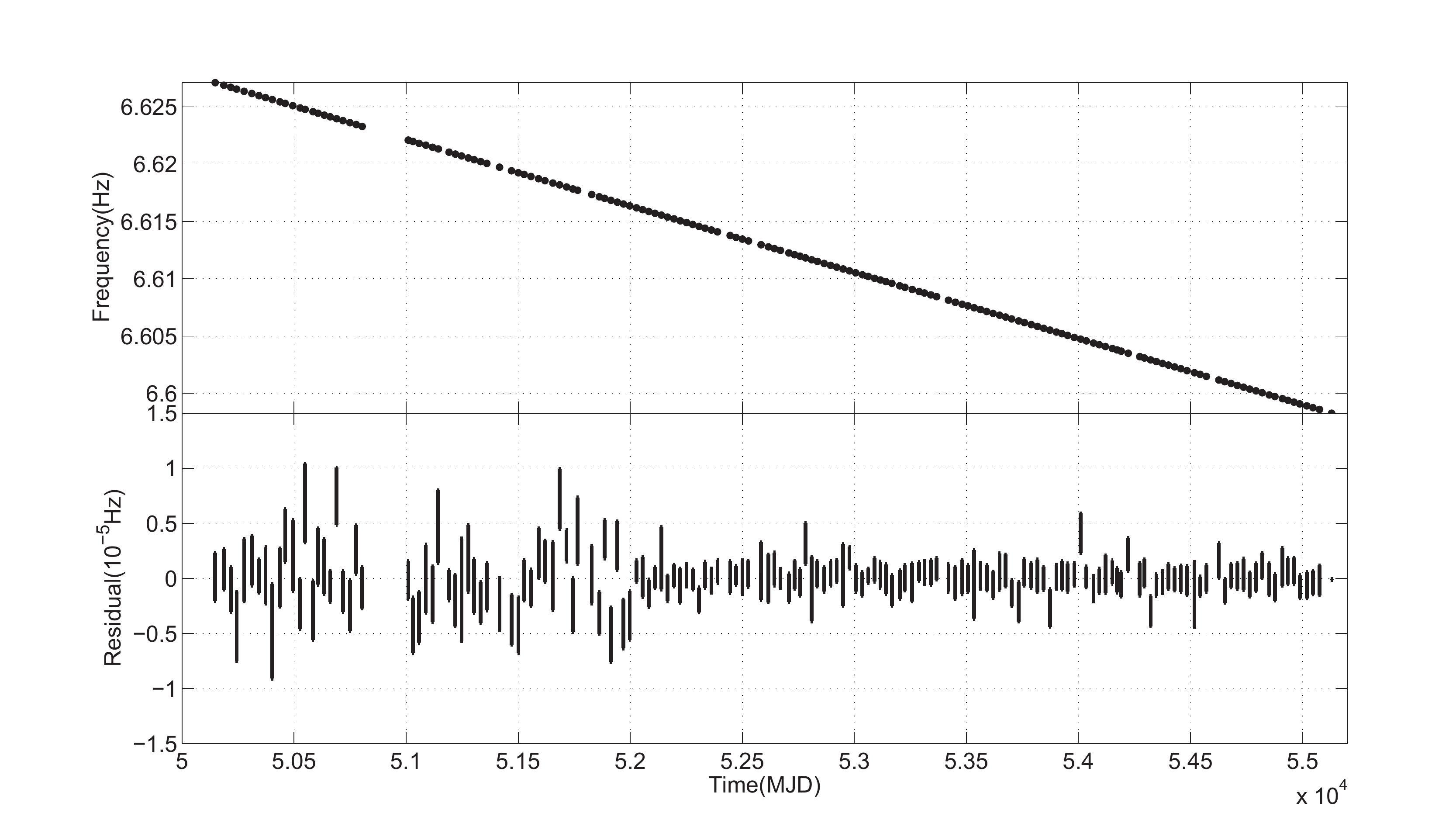}
\caption{The rotation frequencies of PSR B1509-58 obtained from the
X-ray observations (upper pannel) and the residuals of the fit with equation(\ref{eq:1})
(lower pannel). \label{fig2}}

\includegraphics[width=0.8\textwidth]{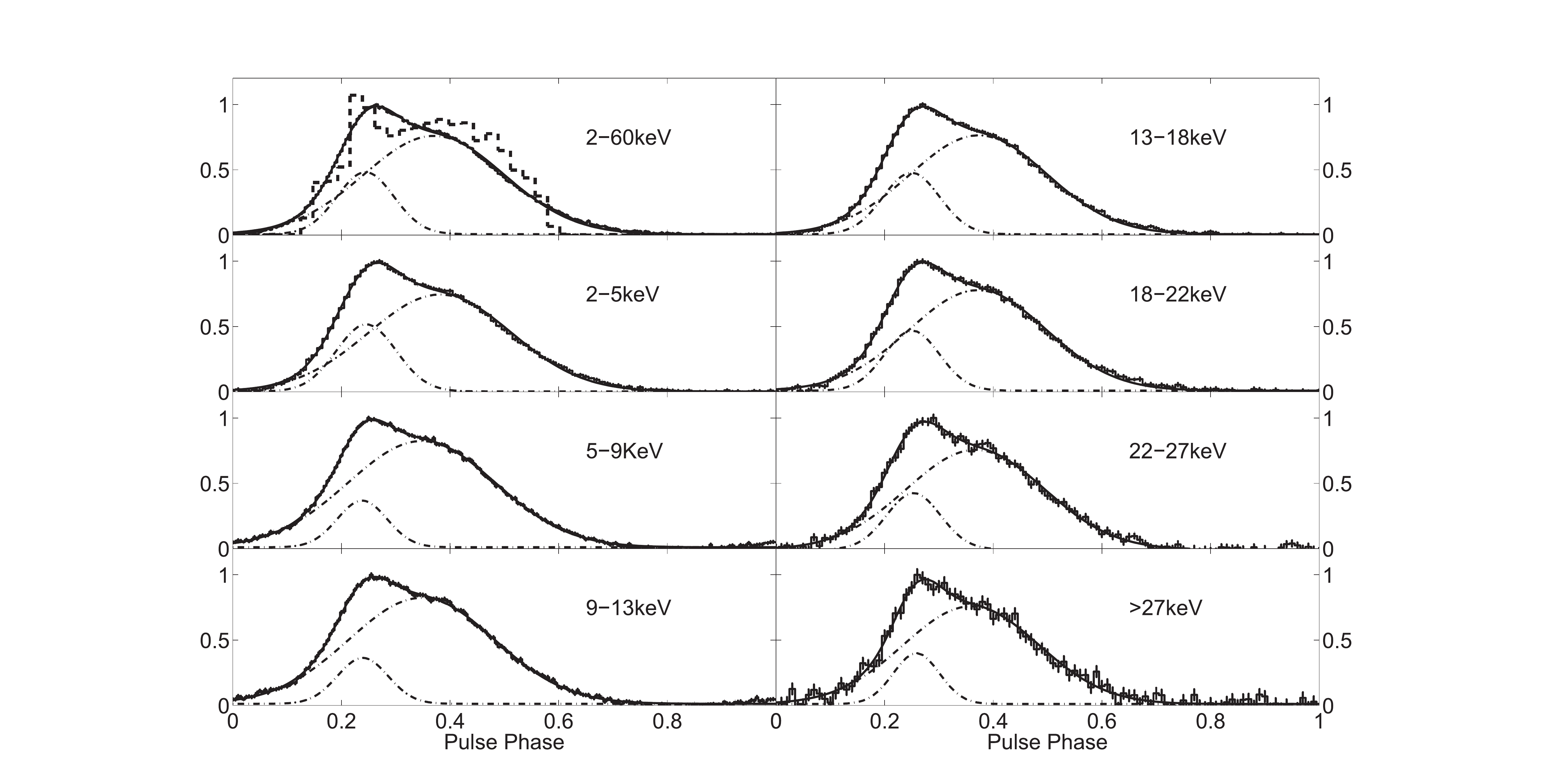}
\caption{The X-ray profiles of PSR B1509-58 in different energy
ranges. The two dot-dashed lines represent the two Gaussian
components used to fit the pulses. The dashed line represents
model result \cite{Zhang and Cheng(2000)}\label{fig7}}

\end{center}
\end{figure*}

For PSR B1509-58 the timing results are showed in Fig.
\ref{fig2}. The timing properties of PSR B1509-58 are
derived in a similar way to that of the Crab pulsar
and the detailed X-ray ephemeris of PSR B1509-58
is listed in Table. \ref{table:2}.

We created the light curves of PSR B1509-58 in the same energy bands
as for the Crab pulsar, while the period of PSR B1509-58 was divided
into 100 or 200 phase bins due to its much lower flux. The X-ray pulse
profile of PSR B1509-58 is broad and asymmetric, with a steep rise
and flat decay, and no evolution with energy could be found by
eye-balls. \cite{Kuiper et al.(1999)} and \cite{ Cusumano et
al.(2001)} found that its X-ray pulse could be well fitted by two
Gaussian functions:
\begin{equation}
L=N_{1}e^{-0.5*(\frac{\phi-\mu_{1}}{\sigma_{1}})^{2}}+N_{2}e^{-0.5*(\frac{\phi-\mu_{2}}{\sigma_{2}})^{2}}+l,
\label{eq:6}
\end{equation}
where $N_{1}$, $\mu_{1}$, $\sigma_{1}$, $N_{2}$, $\mu_{2}$,
$\sigma_{2}$, and $l$ are free parameters.
We therefore fitted the X-ray pulses of PSR B1509-58
in different energy bands with two Gaussian functions so as
to quantify the dependency of the X-ray pulse on photon energy. The
coefficients and the reduced $\chi^{2}$ are listed in Table. \ref{table:B1509profit}.
Fitting the 2-60\,keV profile derives a narrow component peaking at
0.249$\pm$0.001 with a width of 0.053$\pm$0.001 and a broader
component peaking at 0.375$\pm$0.002 with a width of
0.129$\pm$0.002, which are consistent with those obtained by
\citep{Kawai et al.(1991)}, \cite{Rots et al.(1998)}, and \cite{
Cusumano et al.(2001)}. The central phases of the two components are
slightly different from the results of \cite{Kuiper et al.(1999)}
and \cite{Cusumano et al.(2001)}, which may be caused by discrepant
energy bands of the profiles. Fig. \ref{fig7} displays the pulse
profiles in different energy bands fitted with two Gaussian
functions, and Fig. \ref{fig8} plots the relative intensity of the
components as well as their central phases. It could be seen that,
with the increasing photon energy, the second component becomes
stronger and the separation between the two components decreases,
implying that the X-ray pulse of PSR B1509-58 is narrower at higher
energy.

\subsubsection{Phase resolved spectral analyses}

\begin{figure*}
\begin{center}

\includegraphics[width=0.8\textwidth]{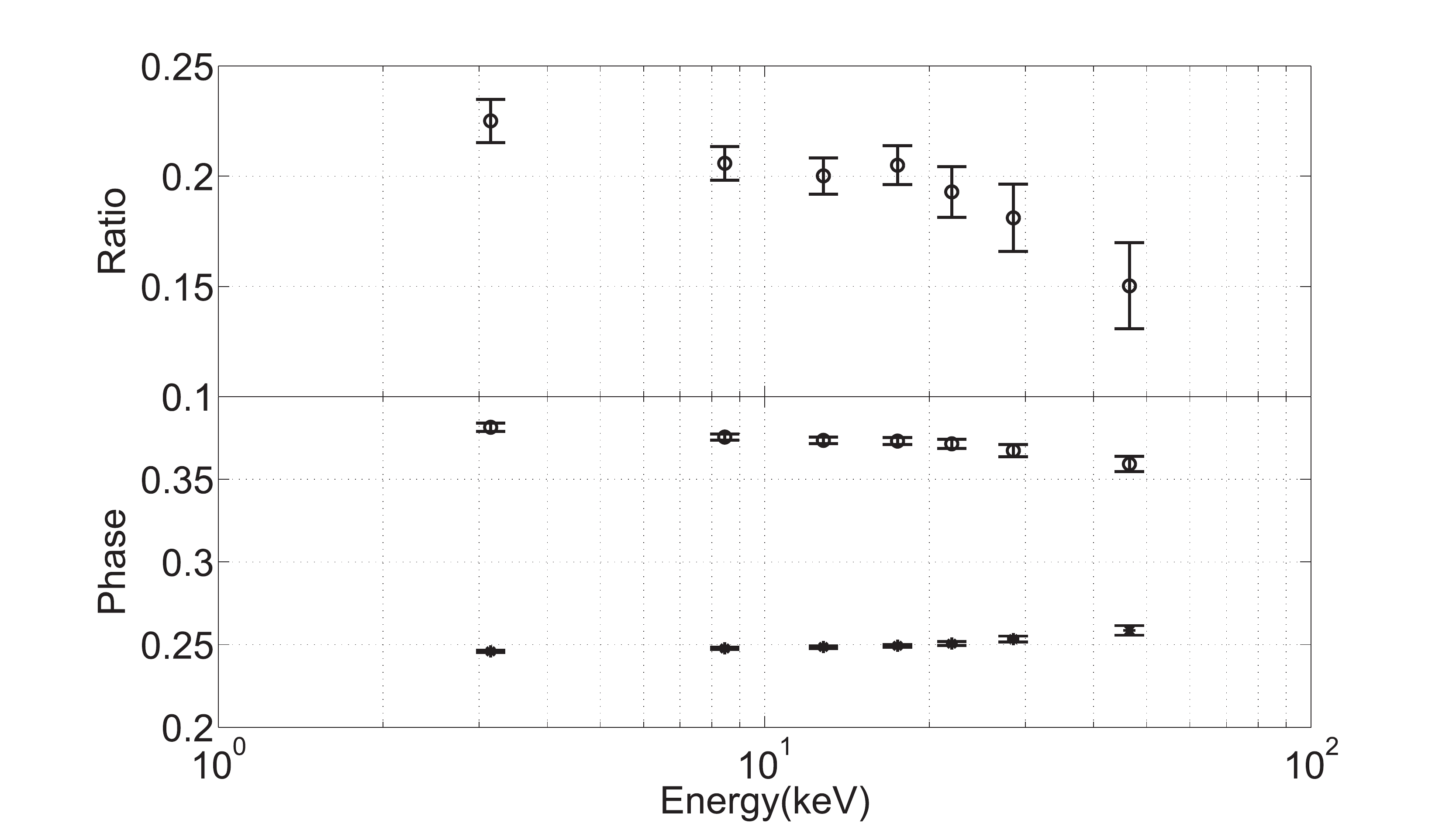}
\caption{Upper pannel: The intensity fraction of the first narrow
Gaussian component of PSR B1509-58 as function of energy. Lower
pannel: the positions of the two components in different energy
bands. The stars and circle points represent the positions of the
narrower and broader components, respectively. \label{fig8}}
\includegraphics[width=0.8\textwidth]{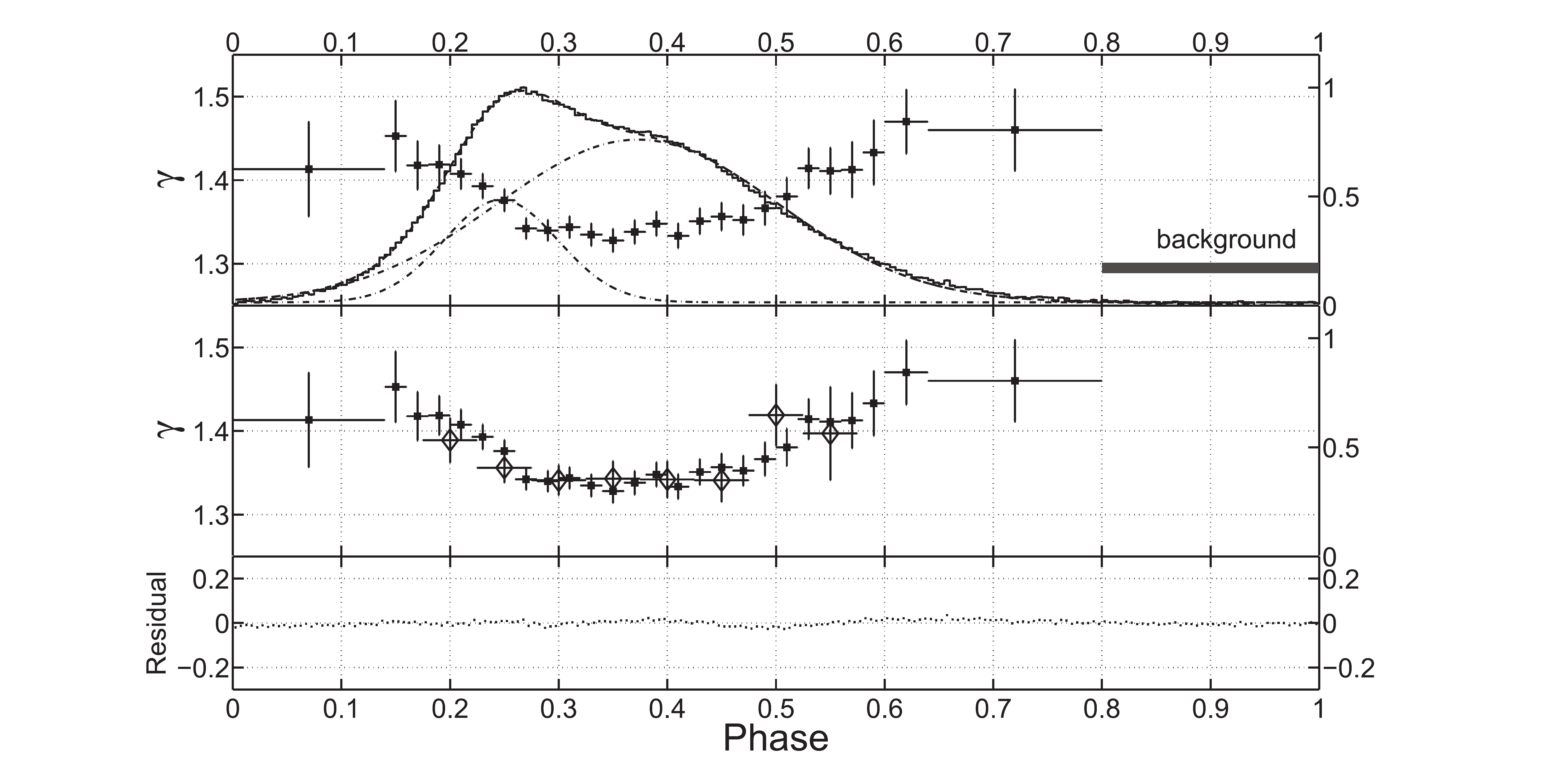}
\caption{Top pannel: The photon indices of PSR B1509-58 in different
phase ranges (``+''). The black belt represents the phase region of
the unpulsed background. Middle pannel: Comparison of the photon
indices obtained in this work and that by \cite{Rots et al.(1998)}.
Bottom pannel: The residuals between the observational profile and
the fitted profile.\label{fig12}}

\end{center}
\end{figure*}

We fitted the X-ray spectrum of the 26 phase bins of PSR B1509-58 by
a single power law with photoelectric absorption in the energy range
3-30\,keV, and the fitting results are listed in Table.
\ref{table:B1509phrespec} and plotted in Fig. \ref{fig12}. In
contrast to the results of Rots et al. (1998), who suggested that
the spectrum does not change with phase, our results show that the
photon index do changes within the pulse. The photon index decreases
from 1.4 at the rising edge of the pulse gradually to about 1.32
just at the maximum phase, then keeps constant, and starts to rise
at the shoulder of the pulse to about 1.45 at the trailing  wing.
Comparing our result to that of Rots et al. (1998) (Fig.
\ref{fig12}, lower panel) shows that these two results are actually
consistent within the error bars, and what makes the difference is
that we have used ten times more data than Rots et al. (1998) and so
get much smaller error bars. If the photon indices
in different phases have the uniform distribution, the $\chi^{2}$ of 26
data points is 335.4, which means photon indices have
high significant evolution with phase. The total absorbed flux of the
pulse is $(2.30\pm0.02)\times$10$^{-11}$\,ergs s $^{-1}$ cm$^{-2}$ in
the 2-10\,keV band, which is consistent with
2.0$\times$10$^{-11}$\,ergs s $^{-1}$ cm$^{-2}$
given by \cite{Cusumano et al.(2001)}.

\subsection{PSR B0540-69}

\subsubsection{Timing and Pulse profiles}
\begin{figure*}
\begin{center}
\includegraphics[width=0.8\textwidth,clip]{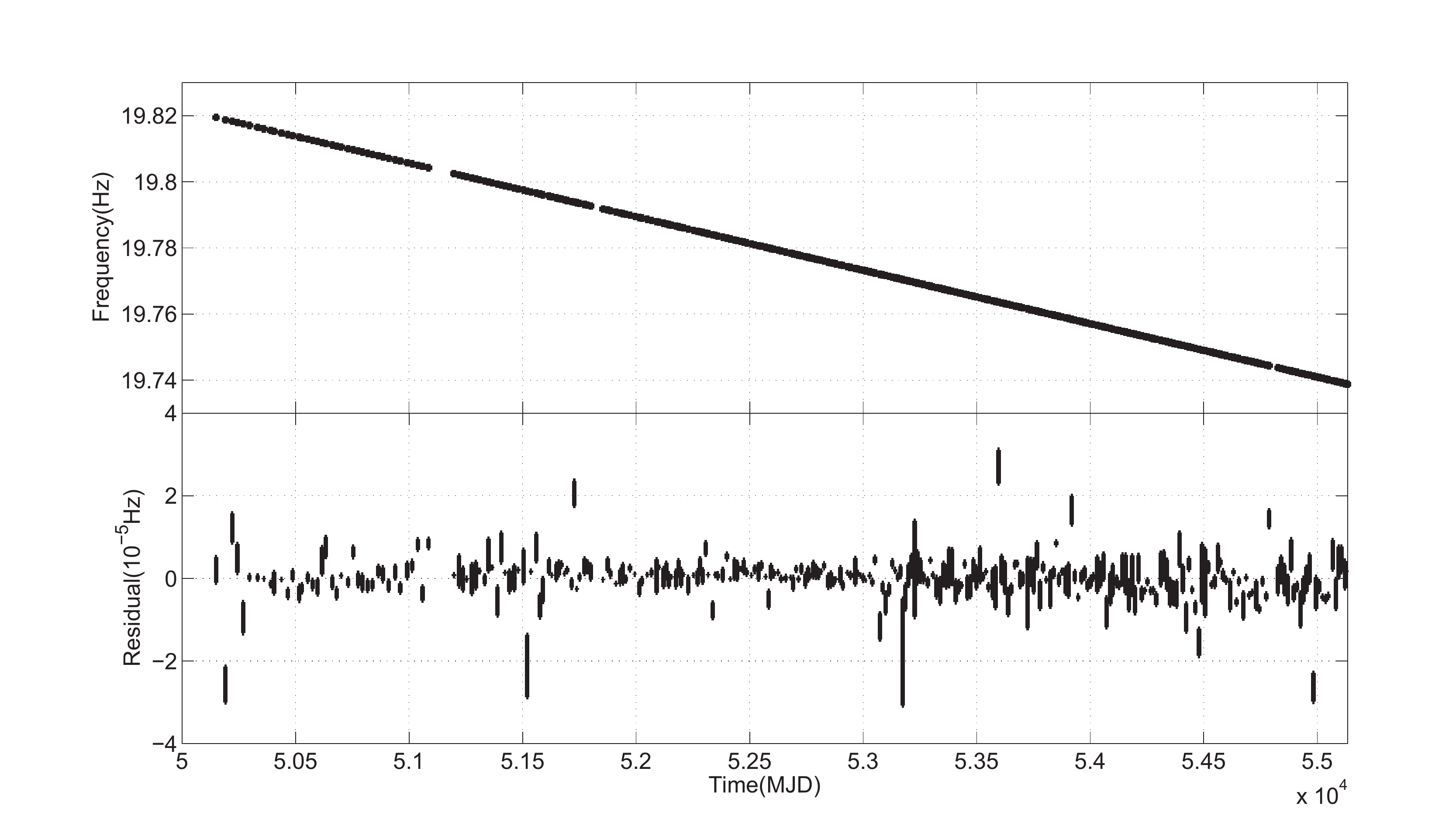}
\caption{Similar to Fig. \ref{fig2} but for PSR B0540-69.\label{fig3}}

\includegraphics[width=0.8\textwidth]{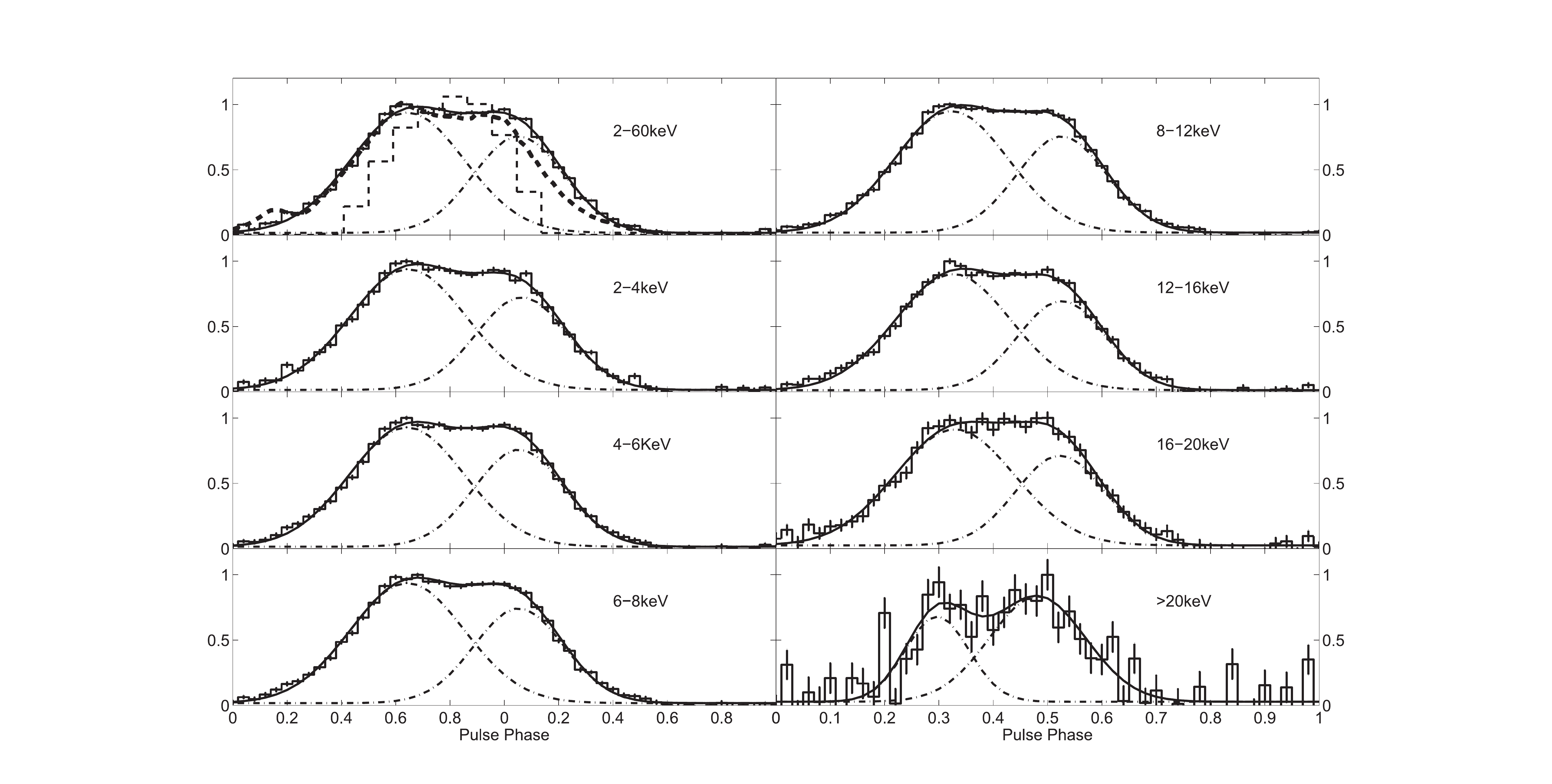}
\caption{The X-ray profiles of PSR B0540-69 in different energy
ranges. The two dot-dashed lines represent the two Gaussian
components used to fit the pulses. The thin and thick dashed lines
represent the model results from \cite{Zhang and Cheng(2000)} and
\cite{Takata and Chang(2007)} separately.\label{fig9}}
\end{center}
\end{figure*}

\begin{figure*}
\begin{center}
\includegraphics[width=0.8\textwidth]{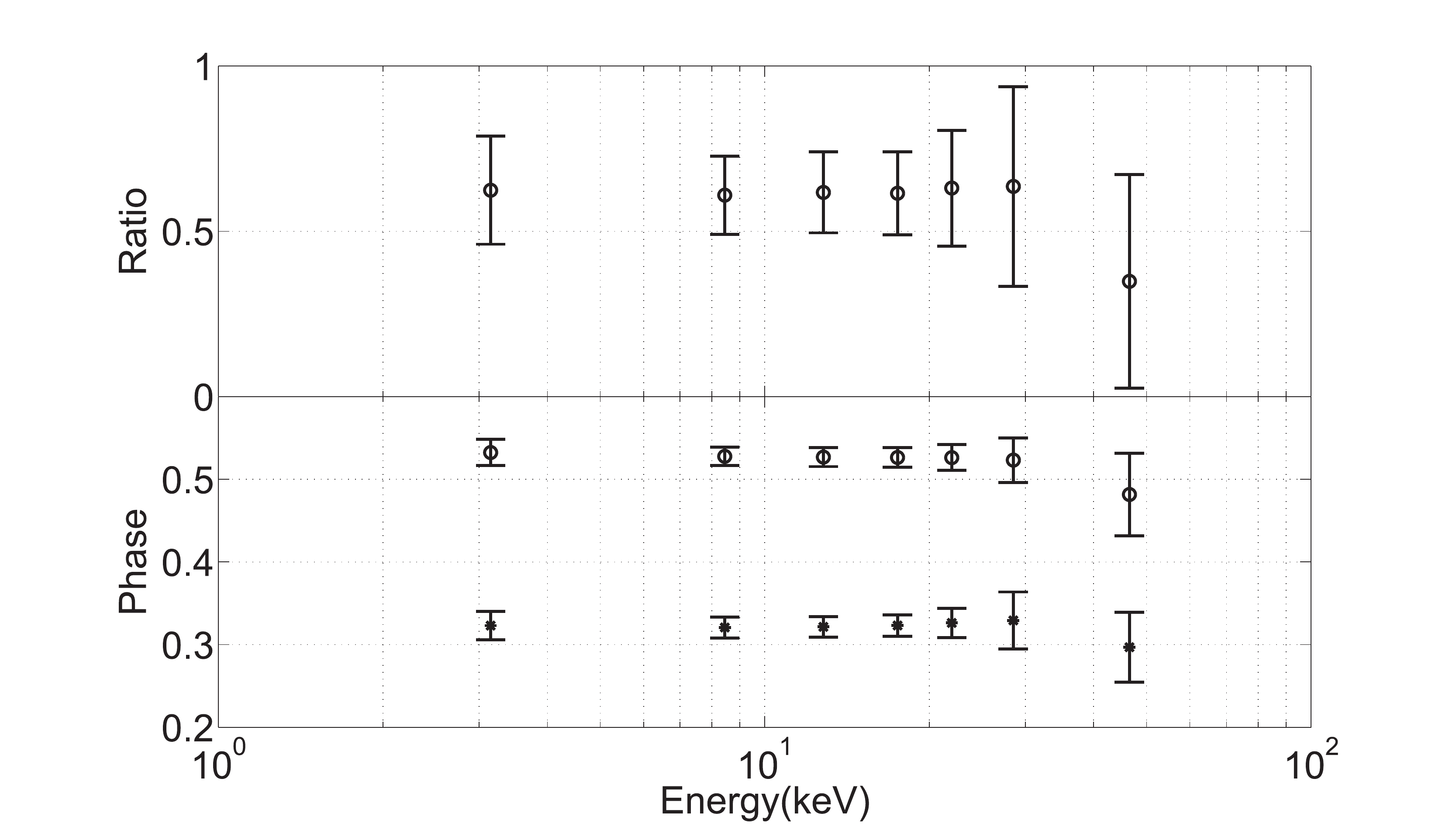}
\caption{Similar to Fig. \ref{fig8} but for PSR B0540-69.\label{fig10}}

\includegraphics[width=0.8\textwidth]{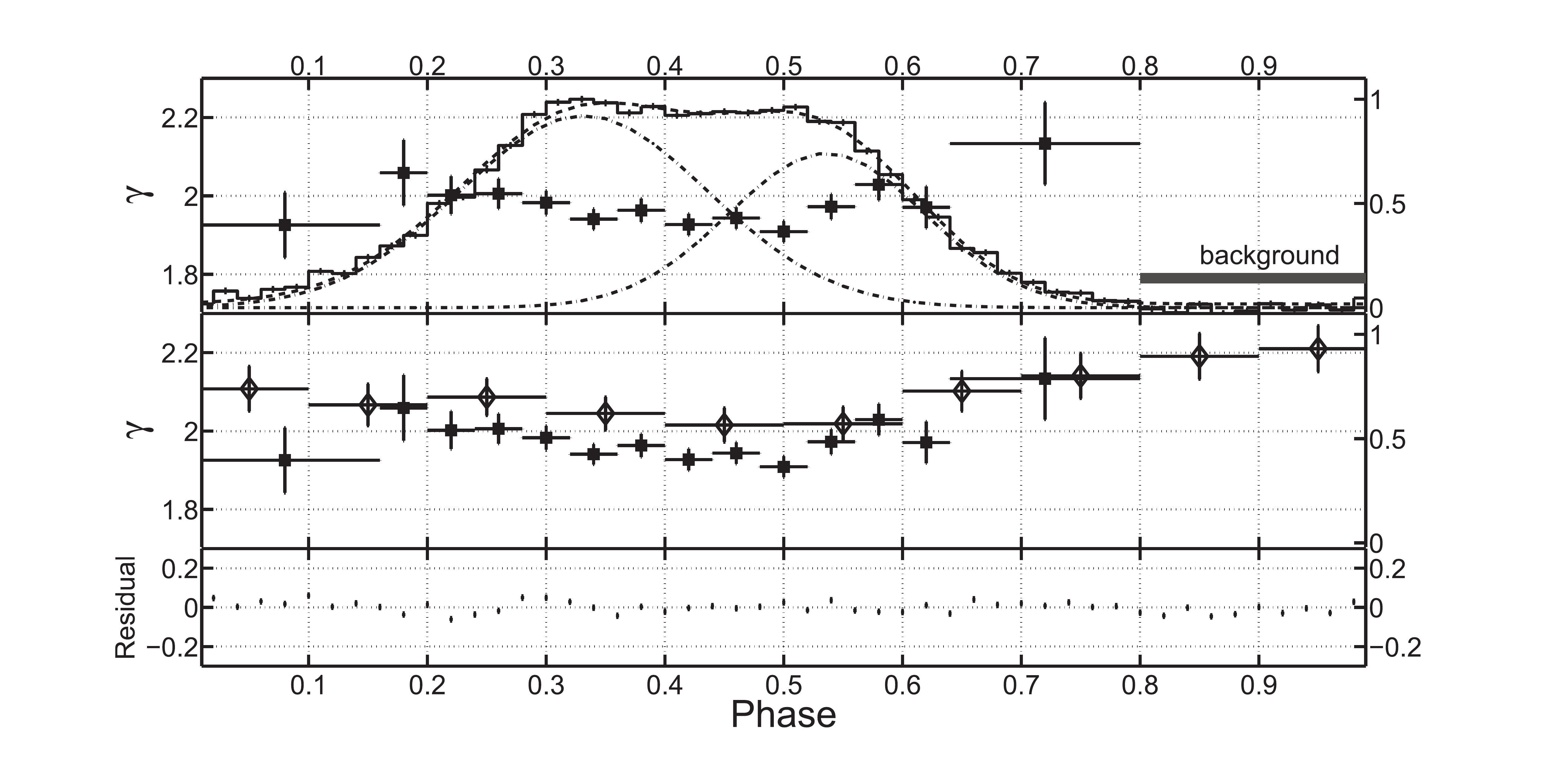}
\caption{Top pannel: The photon indices of PSR B0540-69 in different
phase ranges (``+''). The black belt represents the phase region of
the unpulsed background. Middle pannel: Comparison of the photon
indices obtained in this work and that by \citep{Hirayama et
al.(2002)}. Bottom pannel: The residuals between the observational
profile and the fitted profile.\label{fig13}}

\end{center}
\end{figure*}

For PSR B0540-69, the timing results are showed in Fig. \ref{fig3}.
The timing properties of PSR B0540-69 are derived in
a similar way to that of the Crab pulsar
and the detailed X-ray ephemeris of PSR B0540-69
is listed in Table. \ref{table:2}.

Because PSR B0540-69 is the dimmest among the three pulsars, its
light curves were created in 2-60, 2-4, 4-6, 6-8, 8-12, 12-16,
16-20, and $>$20\,keV bands, correspond to PCA channels 0-255, 0-9,
10-13, 14-18, 19-27, 28-36, 37-47, and 48-255, and one period was
divided into 50 phase bins. PSR B0540-69 has a single broad X-ray
pulse slightly hollowed in the center \citep{Plaa et al.(2003)}. The
pulse profiles with different energy ranges have no significant
variation as from Fig. \ref{fig9}, which is consistent with that
given by \cite{Plaa et al.(2003)} and \cite{Campana et al.(2008)}.
The pulse profiles can also be fitted well with two Gaussian
functions(Equation \ref{eq:6} and Table. \ref{table:B0540profit}).
The central phases of the two functions, except the
profile above 20\,keV that has large uncertainty, are located at
about phase 0.32 and 0.53, and their widths are about 0.11 and 0.08
respectively. The separations between two functions in different
energy ranges are about 0.21 in phase (Fig. \ref{fig9}). These
results are similar to \cite{Plaa et al.(2003)}. The fraction of the
first broader pulse accounting for the whole pulse and the central
phases of the two Gaussian functions are calculated in a similar way
to that for PSR B1509-58. The fraction and the central phases of the
two components do not show significant evolution with energy (Fig.
\ref{fig10}).

\subsubsection{Phase resolved spectral analyses}

Totally we created 14 phase resolved spectra of PSR B0540-69 and
fitted them in 3-30\,keV by a simple power law model with
photoelectric absorption.  The spectral evolution with phase of PSR
B0540-69 is different from the Crab pulsar but similar to PSR
B1509-58 (Table. \ref{table:B0540phrespec} and Fig. \ref{fig13}). The
photon indices in the middle of the pulse (phase range 0.3-0.5) is
about 1.93 but that for the two wings is about 2.0.
The $\chi^{2}$ of 14 data points is 59.8, implying
photon indices of PSR B0540-69 have high significant evolution with phase.
Such a spectral hardening in the center of the pulse was also obtained by
\cite{Hirayama et al.(2002)} using the 1-10\,keV data from the {\sl
ASCA} satellite, though the later suggested a slightly higher photon
index in the central region of the pulse. The photon index of the
whole pulse is 1.92$\pm$0.02. This agrees with the results by
\cite{Mineo et al.(1999)} (1.94$\pm$0.03) and by \cite{Kaaret et
al.(2001)} (1.92$\pm$0.03). The total absorbed flux of the pulse is
(6.4$\pm$0.2)$\times$10$^{-12}$\,ergs s$^{-1}$ cm$^{-2}$, which is
consistent with
6.5$\times$10$^{-12}$\,ergs$\cdot$s$^{-1}\cdot$cm$^{-2}$
\citep{Campana et al.(2008)} in the 2-10\,keV band.

\subsection{Photon Index Evolution with Time}
\begin{figure*}
\begin{center}
\includegraphics[width=0.8\textwidth]{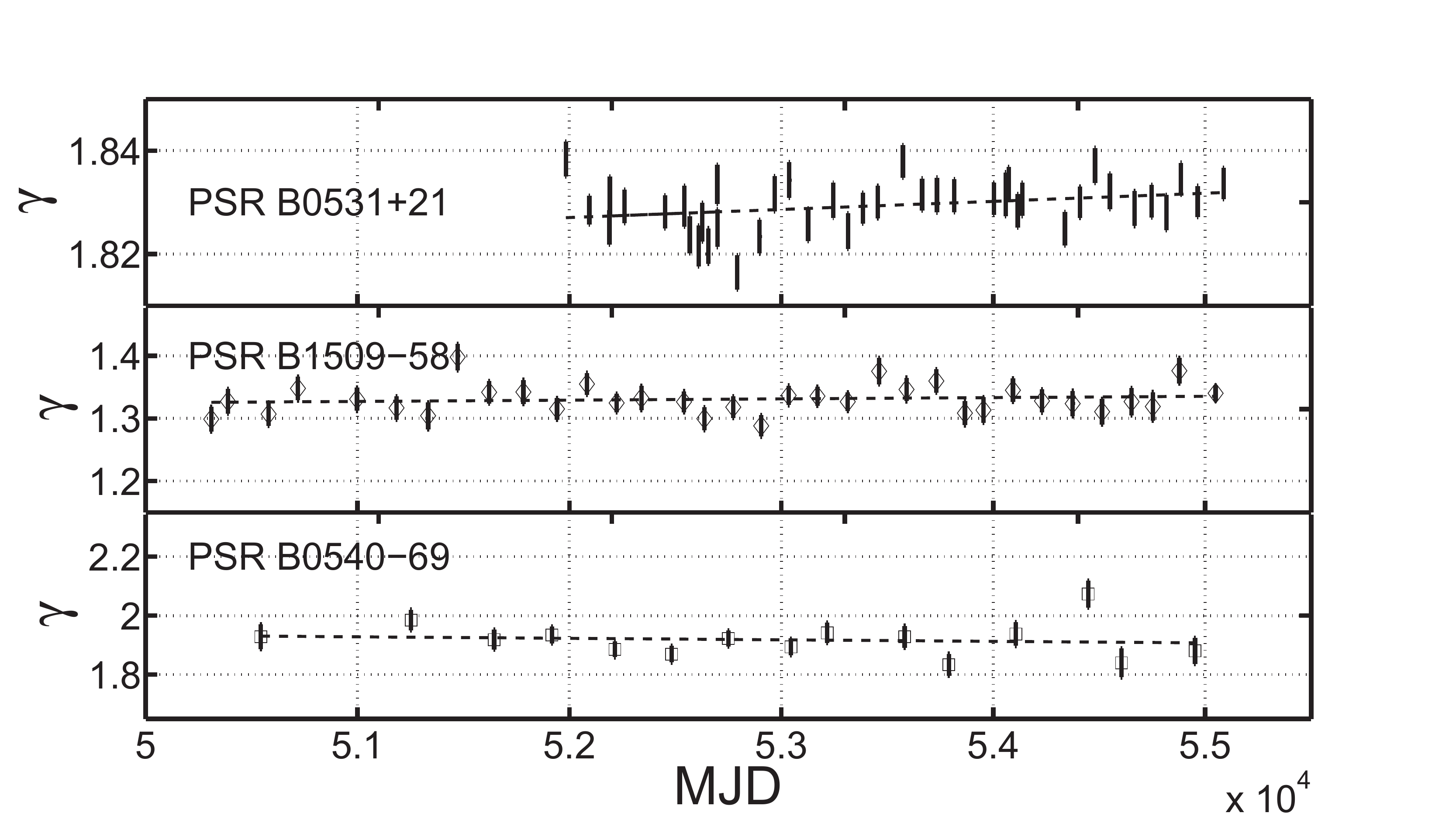}
\caption{The photon indices of PSRs B0531+21, B1509-58, and B0540-69
in different time.\label{fig16}}

\end{center}
\end{figure*}

The long time span of the observations also allows us to study the
real time evolution of the spectra of these three pulsars. Plotting
the photon indices of the whole pulse at different epochs of the
three pulsars suggests that there may exist a slight spectral
softening for the Crab pulsar, while PSR B1509-58 and the spectrum
of PSR B0540-69 remains unchanged (Fig. \ref{fig16}). We note here that the
photon indices of the overall pulsed emission of the Crab pulsar in
an epoch is the mean of the 107 phase resolved photon indices
weighted by the their errors, while those for PSR B1509-58 and
B0540-69 are obtained from the fitting to their whole pulses,
respectively. A linear fit to the Crab data gives a slope of
(1.6$\pm0.7)\times10^{-6}$ day$^{-1}$, that for PSR B1509-59 is
(2.0$\pm3.6)\times10^{-6}$day$^{-1}$, and
(-5$\pm14)\times10^{-6}$day$^{-1}$ for PSR B0540-69. Statistically
only the Crab pulsar has a significant spectral softening. We note
that \cite{Hirayama et al.(2002)} detected no spectral variation of
PSR B0540-69 either.

\begin{figure*}
\begin{center}
\includegraphics[width=0.8\textwidth]{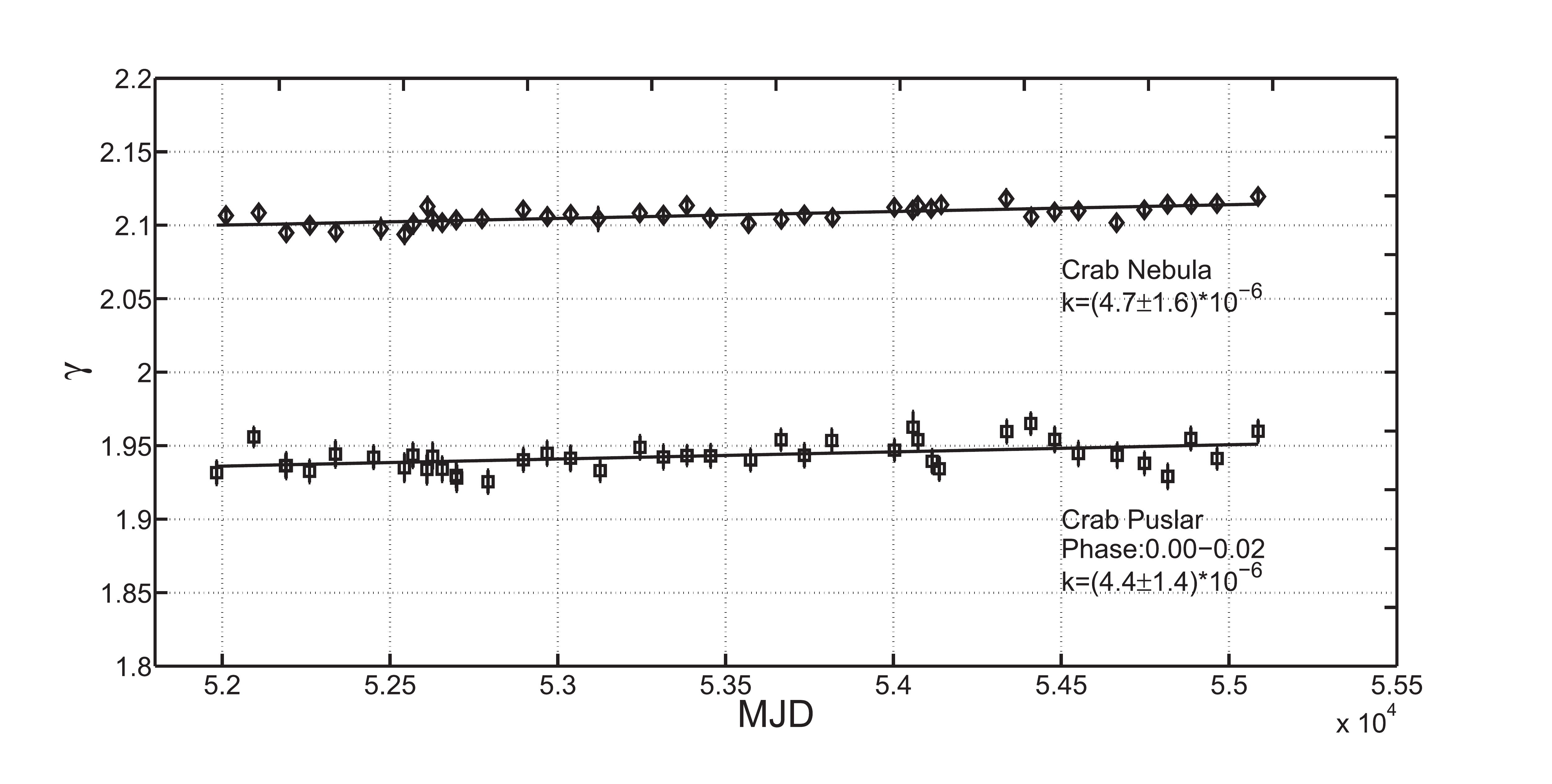}
\caption{The photon indices of the Crab pulsar and the Crab Nebula.
The black lines are the linear fits
to the data and $k$ is the value of the slope.\label{fig15}}

\end{center}
\end{figure*}

However, the Crab nebula including the pulsar is the calibration
source of {\sl {\sl RXTE}}. The spectral
softening is probably a consequence of the improper calibration. In
order to check whether the spectral evolution results we obtained
above is reliable or not, we first studied such spectral evolution
for different phase ranges, which is carried out under the
consideration that the improper calibration would make the spectra
in different phase-bins to have the similar softening slope. We find
that all the slopes are higher than zero, and there do exists a
structure in the phase range of the main peak, i.e.,
the spectral softening is more significant in the
two wings. However, the $\chi^2$ is 24.4(18 d.o.f), showing that the
spectral variety in each phase has the similar softening slope.
On another hand, the spectra of the Crab Nebula were also
analysed with the absorbed power law model and were
extracted from the Standard 2 data of PCU2 with all layers in epoch
5 of {\sl RXTE}. The spectral indices were fitted with a linear
function and the slope is $(4.7\pm1.6)\times10^{-6}day^{-1}$, which
is comparable to the mean slope of all phases: $(5.3\pm0.5)\times10^{-6}day^{-1}$
(Fig. \ref{fig15}). For the rest PCUs, the results are similarly with
\cite{Jahoda et al.(2006)} and \cite{Weisskopf et al.(2010)}.
The spectrum in every phase softening with time may be caused
by the tiny deviation of the calibration.

\section{Discussion}

\subsection{Comparison of the pulse profiles with model predictions}

The X-ray pulse of the Crab pulsar is composed of two peaks and
inter peak region and changes significantly with energy. \cite{Jia
et al.(2007)} have calculated the light curves of the Crab
pulsar with assumption that inner boundary is extended inward
from the null charge surface to 10 stellar radii(Fig. \ref{fig4}).
The X-ray profile of the Crab pulsar predicted by their model
is consistent with the observational data except the second
peak with inclination angle 50$^{\textordmasculine }$ and
viewing angle 75$^{\textordmasculine }$. The intensity
of the leading edge of the second peak is evidently
smaller than the observational profile. \cite{Tang et al.(2008)}
gave a profile in Fig. 5 of their paper. However, that profile is
obviously different from our results. Specifically, the intensity of the
second peak in their paper is higher than that of the first peak,
while in our observation the first peak has significantly higher
intensity. The profile predicted by \cite{Harding et al.(2008)} has at
least 4 peaks, which has not be detected in our high statistical results,
and thus their model is unsuitable. \cite{Du et al.(2010)} only give the high
energy pulse profiles of the Crab pulsar($>$0.1GeV), while our observation
is in the energy range lower than 60 keV.

The X-ray pulses of PSRs B1509-58 and B0540-69 are
broad asymmetric profiles showing no obvious change with energy,
which are different from the Crab pulsar. The X-ray pulses of
PSRs B1509-58 and B0540-69 can be fitted with two Gaussian
curves\citep{Kuiper et al.(1999), Cusumano et al.(2001), Plaa et al.(2003)}.
\cite{Zhang and Cheng(2000)} have calculated the
light curves of PSRs B1509-58, and B0540-69
(Fig. \ref{fig7}, and Fig. \ref{fig9}). For PSR B1509-58,
the predicted profile is sharp at edges and the
top of profile has two peaks comparing with the observed profile
with inclination angle 60$^{\textordmasculine }$ and viewing angle
75$^{\textordmasculine }$. For PSR B0540-69, the predicted profile
is narrower than the observed profile with inclination angle
50$^{\textordmasculine }$ and viewing angle 76$^{\textordmasculine
}$. \cite{Takata and Chang(2007)} recalculate the profile of PSR
B0540-69 (Fig. \ref{fig9}) with inclination angle
30$^{\textordmasculine }$ and viewing angle 81.5$^{\textordmasculine
}$. The profile of PSR B0540-69 predicted by their model are
consistent with the observational results. The small peak located
at the leading edge of the pulse is larger while the trailing
edge of pulse is smaller than the observation profile. However,
these models give surprising results to interpret the observations,
more specifics need to be considered in detailed modeling.

\subsection{Constrains from the phase-resolved spectra}

\begin{figure*}
\begin{center}

\includegraphics[width=0.8\textwidth]{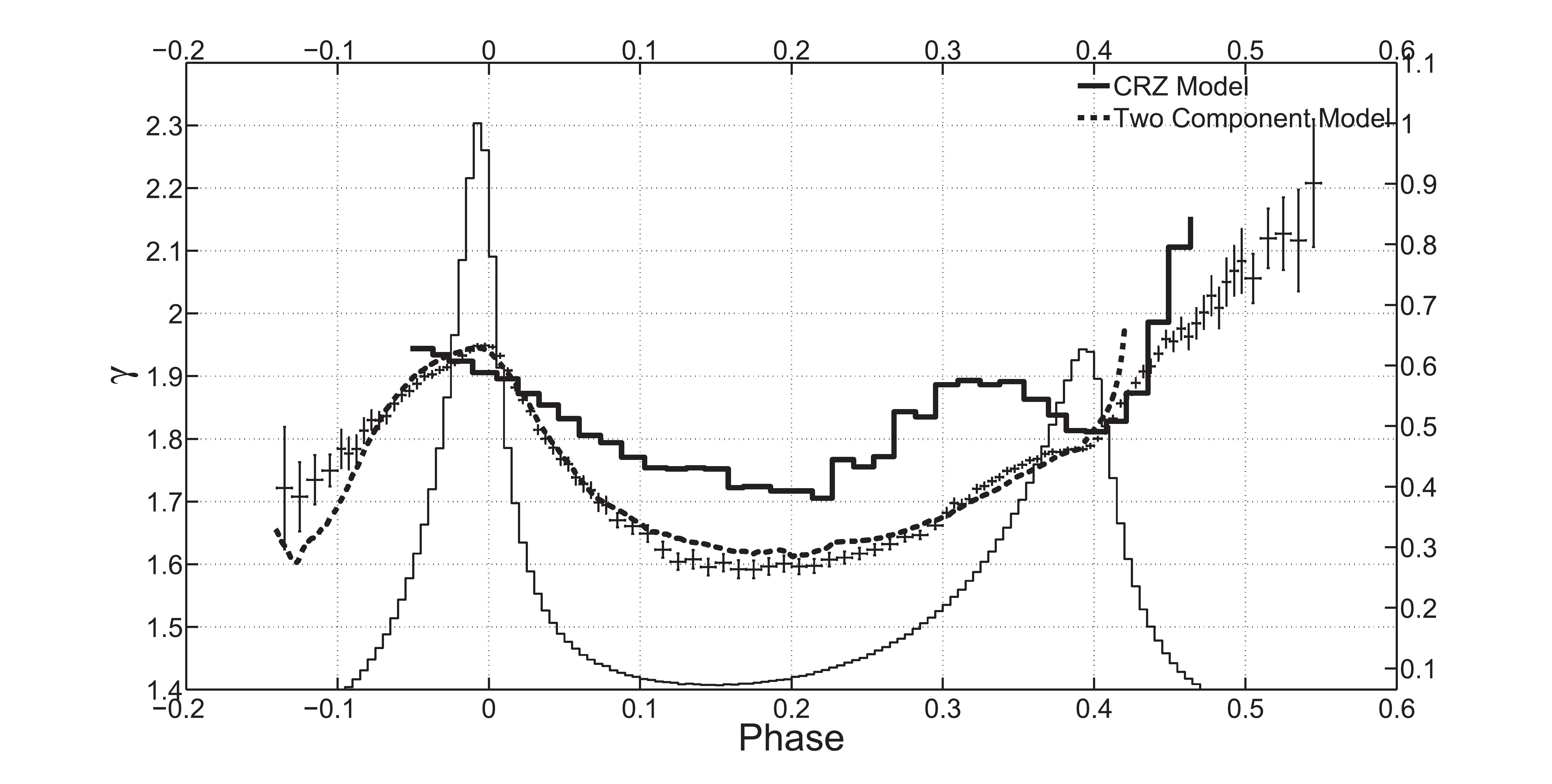}
\caption{The photon indices of the Crab pulsar in different phase ranges (``+''). The thick
solid line represents the model prediction by the CRZ model \citep{Zhang and Cheng(2002)}
and the dashed line represents the two-component model\citep{Massaro et al.(2000)}.
\label{fig113}}

\end{center}
\end{figure*}

We confirmed the variation of Crab's spectrum with phase as detected
in the previous studies, and found a rapid softening in the trailing
edge of the second peak. \cite{Pravdo et al.(1997)} and \cite{Massaro et al.(2000)}
have given the similar results, which are plotted in Fig. \ref{fig114}.
Comparing to their result, we have obtained more spectral
indices, including the leading edge of the first peak and the
trailing edge of the second peak. Our results show that the photon
index variation between the two peaks is very smooth, without any
small bump that can not be excluded by the previous low statistic
analyses (e.g., Pravdo et al. 1997 and Massaro et al. 2000).
We find that hardest spectrum occurs positionally coincide
with the minimum intensity in the inter
peak region, which is unclear in the analyses of \cite{Pravdo et
al.(1997)}. Those new results set stronger constraints on the
theoretical models of pulsar high energy emission. \cite{Zhang and
Cheng(2002)} calculated the phase resolved X-ray spectrum of the
Crab pulsar using the CRZ model with only the synchrotron radiation
from the secondary pairs being considered. Their calculation can
explain the rapid spectral softening of the trailing edge of the
second peak (Fig. \ref{fig113}). However, from the comparison we know
that their model prediction is different from observations in the
following aspects: (1) the observed spectrum in the leading edge of
the first peak softens when approaching the intensity maximum and
then hardens, while the model prediction is that the spectrum
hardens all through the first peak; (2) the spectrum of the
inter-peak region is significantly and systematically harder than
the model prediction; and (3) the model predicts ahead of the second
peak a softer region that can be excluded by observations. These
discrepancies will set tight constrains on the future pulsar X-ray
emission models and their parameters.

The phase resolved spectra of PSR B1509-58  are calculated here in
narrower and more phase bins compared to the work of \cite{Rots et
al.(1998)} (diamond points in Fig. \ref{fig12}). They vary with
phase and do not follow the intensity of the profile. The photon
indices are about 1.33$\pm$0.01 in the center part of the pulse
(0.25-0.5) and become softer in the wings. Such a spectral evolution
is apparently different from the behaviors of the two peaks of the
Crab pulsar. However, if the pulse profile of PSR B1509-58 is a
combination of two Gaussian components as pointed out by
\cite{Kuiper et al.(1999)} and \cite{Cusumano et al.(2001)}, the
hardest spectrum will be between the two components (Fig. \ref{fig7}
and Fig. \ref{fig12}), which is similar to that of the Crab pulsar.
The phase resolved spectra of PSR B0540-69 are similar to PSR
B1509-58, albeit with lower statistics and thus larger
uncertainties. The broad pulse of PSR B0540-69 can be de-composed
into two narrower Gaussian pulses too \citep{Plaa et al.(2003)}.
Thus its overall spectral evolutionary trend is also similar to that
of the Crab pulsar, as shown in Fig. \ref{fig13}. \cite{Massaro et al.(2000)}
proposed that the pulsed emission of the Crab pulsar
can be described as the superposition of optical component($C_{O}$) and
X-ray component($C_{X}$). The phase-resolved spectra of the Crab
pulsar in the two-component model can also be calculated
from the ratio of the summed intensities of $C_{O}$ and $C_{X}$ at two extreme values:
\begin{equation}
\gamma_{12}=\frac{Log(F(E_{2})/F(E_{1}))}{Log(E_{2}/E_{1})}+\gamma_{O},
\label{eq:13}
\end{equation}
where $E_{1}=3$\,keV, $E_{2}=60$\,keV, and $\gamma_{O}=1.95$.
As shown in Fig. \ref{fig113}, the results agree with the spectral
evolution of the Crab pulsar except at the tail edge of
the second peak, where the model predicted photon indices
are smaller than the observed ones.

\section{Summary}

It is shown with the {\sl {\sl RXTE}} data that the spectra of PSRs
B0531+21, B1509-58 and B0540-69 all evolve with pulse phase. In
addition to the previous results of PSR B0531+21, which has been
confirmed with a much higher statistics, we found that the spectral
softening increases immediately after the second peak. Our
analyses show that the spectra of PSR B1509-58 and B0540-69 are
harder at the centers of the pulses and are softer at the
wings, which are different from that of the Crab pulsar, whose
spectrum is softer at the two peaks. However, since the pulses of
PSRs B1509-58 and B0540-69 could be further decomposed each into two
narrower Gaussian components, their overall spectral evolution with
phase is thus similar to the Crab pulsar. These new results set more
constraints on pulsar X-ray emission models.

\section*{Acknowledgments}
We thank the referee for his/her insightful and very useful comments
and suggestions.  Drs. Hao Tong, Shanshan Weng and Xiaobo Li
also helped us a lot in the preparation of the manuscript. We thank
the High Energy Astrophysics Science Archive Research Center (HEASARC) at
NASA/Goddard Space Flight Center for maintaining its online archive
service which provided the data used in this research. This work is
supported by National Basic Research Program of China (973 program,
2009CB824800) and National Science Foundation of China(11133002).

\clearpage
\begin{table*}

\footnotesize
%%\tiny
\caption{The parameters of PSRs B0531+21, B1509-58 \& B0540-69}
\scriptsize{} \label{table:pulsarpar}
\medskip
\begin{center}
\begin{tabular}{l l l l l l l l l l l}
\hline\hline
    & Pulsar Name  &  P & $\dot{P}$  &  $\tau$ & $B$ &$E_{rot}$& $L_{X}$ & $\gamma$ & Braking Index\\
    & & (ms) & $(10^{-15}\,ss^{-1})$ & (yr) & $(10^{12} G)$ & $(10^{38}\,erg s^{-1})$ & $(10^{36}\,erg s^{-1})$ & & &\\
\hline
    & Crab &  33  &  423  &  1240 & 1.2 & 4.7 &  1.3$^{a}$  &  2.022(14)$^{1}$  &  2.509(1)$^{8}$  &\\
\hline
    & PSR B1509-58 &  150  &  1540  &  1560  & 4.8 & 0.18 &  0.47$^{b}$  &  1.358(14)$^{2}$  &  2.837(1)$^{9}$  &\\
    & & & & & & & & & 2.80(3)$^{10}$ &\\
\hline
    & PSR B0540-69 &  50  &  480  &  1670  & 1.6 & 1.5 &  4.0$^{a}$  &  2.1(2)$^{3}$  &  1.81(7)$^{11}$  &\\
    & & & & & & & & 1.845(4)$^{4}$ & 2.125(1))$^{12}$ &\\
    & & & & & & & & 1.83(13)$^{5}$ & 2.140(9)$^{13}$ &\\
    & & & & & & & & 1.94(3)$^{6}$ &  &\\
    & & & & & & & & 1.3(5)$^{7}$ &  &\\

\hline
\end{tabular}
\end{center}

Note: 1) 1: \cite{Kuiper et al.(2001)}, 2: \cite{Marsden et
al.(1997)}, 3: \cite{Campana et al.(2008)}, 4: \cite{Plaa et
al.(2003)}, 5: \cite{Kaaret et al.(2001)}, 6: \cite{Mineo et
al.(1999)}, 7: \cite{Finley(1993)}, 8: \cite{Lyne et al.(1988)}, 9:
\cite{Kaspi et al.(1994)}, 10: \cite{Simon Johnston and David
Galloway (1999)}, 10: \cite{Zhang et al.(2001)}, 12: \cite{Cusumano
et al.(2003)}, 13: \cite{Livingstone et al.(2005)}

2) a: energy range: 2-10\,keV (\cite{Kaaret et al.(2001)}) b: energy
range: 2-250\,keV (\cite{Marsden et al.(1997)})

\footnotesize
%%\tiny
\caption{The timing results of PSRs B0531+21, B1509-58, and B0540-69 from the X-ray
Observations} \scriptsize{} \label{table:2}
\medskip
\begin{center}
\begin{tabular}{l l l l l}
\hline\hline
    &   & PSR B0531+21  &  PSR B1509-58 & PSR B0540-69  \\
\hline
    & R.A. & $05^{h}34^{m}31^{s}.972$ &$15^{h}13^{m}55^{s}.620$ &$05^{h}40^{m}11^{s}.202$\\
    & Decl. & 22$^{\textordmasculine
}$00$^{\prime}$52$^{\prime\prime}$.07 & -59$^{\textordmasculine
}$08$^{\prime}$9$^{\prime\prime}$.00 & -69$^{\textordmasculine
}$19$^{\prime}$54$^{\prime\prime}$.17\\
    & Epoch(MJD) & 52247.0 &52656.0 & 52656.0\\
    & F0(Hz) & 29.82341787(28) & 6.61254815(23) & 19.7787959(11)\\
    & $\dot{F}$(10$^{-10}$s$^{-2}$) & -3.740484(16) & -0.669734(13) & -1.873520(65)\\
    & $\ddot{F}$(10$^{-21}$s$^{-3})$ & 9.545(24) & 1.933(22) & 3.73(11)\\
    & Braking index & 2.035(5) & 2.85(3) & 2.10(6)\\

\hline
\end{tabular}
\end{center}
\end{table*}

\begin{table*}

\footnotesize
%%\tiny
\caption {The list of {\sl RXTE} observations used in this work}
\scriptsize{}\label{table:ListObsID}
\medskip
\begin{center}
\begin{tabular}{ c c c c c c c}
\hline \hline

& Pulsar & Obs ID & Start Date & End Date & offset(') & exposure(s) \\
\hline

& & 50099 & 2001-02-15 & 2001-08-27 & 0.03 & 7504 \\

& & 60079 & 2001-09-10 & 2002-10-22 & 0.03 & 20240 \\

& & 60080 & 2001-07-18 & 2001-07-20 & 0.03 & 3776 \\

& & 60420 & 2001-09-07 & 2001-09-09 & 0.03 & 1824 \\

& & 70018 & 2002-05-09 & 2003-05-14 & 0.03 & 9616 \\

& Crab & 70802 & 2002-11-07 & 2003-02-26 & 0.03 & 7888 \\

& & 80802 & 2003-03-13 & 2004-02-15 & 0.03 & 17552 \\

& & 90802 & 2004-02-29 & 2005-02-25 & 0.03 & 18032 \\

& & 91802 & 2005-03-13 & 2006-02-10 & 0.03 & 17088 \\

& & 92802 & 2006-03-11 & 2006-09-24 & 0.03 & 23536 \\

& & 93802 & 2007-07-17 & 2008-12-17 & 0.03 & 28768 \\

& & 94802 & 2008-12-31 & 2009-11-07 & 0.03 & 16768 \\

\hline

& & total exposure(s) & & & & 172592  \\

\hline

& & 10208 & 1996-03-06 & 1996-10-16 & 0.01 & 14640 \\

& & 20802 & 1996-11-15 & 1997-12-21 & 0.01 & 39056 \\

& & 30704 & 1998-07-15 & 1999-01-13 & 0.01 & 17984 \\

& & 40704 & 1999-02-10 & 2000-03-16 & 0.01 & 43920 \\

& & 50705 & 2000-04-20 & 2001-03-29 & 0.01 & 39408 \\

& PSR B1509-58 & 60703 & 2001-04-28 & 2002-03-28 & 0.01 & 53520 \\

& & 70701 & 2002-04-25 & 2003-04-04 & 0.01 & 51056 \\

& & 80803 & 2003-05-22 & 2004-02-26 & 0.01 & 45408 \\

& & 90803 & 2004-03-25 & 2005-02-18 & 0.01 & 52080 \\

& & 91803 & 2005-03-21 & 2006-02-23 & 0.01 &  48464 \\

& & 92803 & 2006-03-22 & 2007-06-22 & 0.01 & 75744 \\

& & 93803 & 2007-07-12 & 2008-12-19 & 0.01 & 80256 \\

& & 94803 & 2009-01-20 & 2009-10-26 & 0.01 & 44160 \\

\hline

& & total exposure(s) & & & & 578288  \\

\hline

& & 10206 & 1996-08-11 & 1996-11-17 & 0.01 & 11088 \\

& & 10218 & 1996-10-12 & 1996-12-22 & 21.7 & 31552 \\

& & 10250 & 1996-10-04 & 1996-03-08 & 24.9 & 14016 \\

& & 20188 & 1996-12-30 & 1997-12-12 & 24.8 & 51744 \\

& & 30087 & 1998-01-04 & 1998-09-30 & 24.8 & 40624 \\

& & 40139 & 1999-01-19 & 2000-02-15 & 16.1 & 187392 \\

& & 50103 & 2001-03-15 & 2001-03-05 & 15.9 & 246880 \\

& & 50414 & 2000-06-05 & 2000-06-23 & 5.3 & 7040 \\

& PSR B0540-69 & 60082 & 2001-03-26 & 2003-02-21 & 15.8 & 265936 \\

& & 70092 & 2002-03-21 & 2003-04-04 & 15.9 & 292032 \\

& & 80089 & 2003-04-16 & 2004-10-29 & 15.9 & 339904 \\

& & 80118 & 2004-01-07 & 2004-01-09 & 24.4 & 48064 \\

& & 90075 & 2004-03-10 & 2005-05-09 & 16.0 & 188256 \\

& & 91060 & 2005-05-24 & 2006-05-14 & 15.9 & 216496 \\

& & 92010 & 2006-03-12 & 2007-06-22 & 15.9 & 250976 \\

& & 93023 & 2007-07-02 & 2007-06-30 & 15.9 & 304192 \\

& & 93448 & 2008-09-09 & 2008-11-17 & 15.9 & 39296 \\

& & 94023 & 2008-12-26 & 2009-10-28 & 15.9 & 165392 \\
\hline

& & total exposure(s) & & & & 2662528  \\

\hline
\end{tabular}
\end{center}

\end{table*}

\begin{table*}

\footnotesize
%%\tiny
\caption {The coefficients of the Nelson's formula for the main peak and the second peak of the Crab pulsar}
\scriptsize{}\label{table:nelsonpar}
\medskip
\begin{center}
\begin{tabular}{l c c c c c c c c c}
\hline \hline
&& Energy range(keV) & a & b & c & d & e & $\chi^{2}_{dof}$ & d.o.f.\\
\hline

&           &2-60&  -12.33193 & 1078.13826 &   15.05149 & 3870.43735 &  195.75749 & 2.8 & 342 \\
&           &2-5&  -13.76710 & 1001.72137 &    6.87700 & 3843.36216 &  203.31147 & 2.0 & 342 \\
&           &5-9&  -11.79666 & 1100.25133 &   16.59565 & 3858.08309 &  191.78127 & 1.8 & 342 \\
& Main peak &9-13&   -7.75976 & 1069.76546 &   27.38679 & 3535.79500 &  177.79887 & 1.6 & 342 \\
&           &13-17&   -5.96624 & 1205.02963 &   33.87094 & 3629.01368 &  173.93670 & 1.4 & 342 \\
&           &17-22&   -3.54120 & 1102.97976 &   35.51093 & 3210.66333 &  167.27991 & 1.4 & 342 \\
&           &22-27&   -0.18720 & 1202.12570 &   42.07548 & 3268.22696 &  160.77189 & 1.4 & 342 \\
&           &27-60&    0.86096 & 1527.74475 &   33.16336 & 4033.16928 &  174.22816 & 1.1 & 342 \\

\hline

&             &2-60&  -23.85914 &  259.29529 &  -37.77808 &  516.20914 &   97.90285 & 2.9 & 443 \\
&             &2-5&  -23.97517 &  280.84606 &  -38.68251 &  577.13359 &   98.38800 & 1.8 & 443 \\
&             &5-9&  -23.76023 &  251.13817 &  -37.33119 &  495.12274 &   97.12726 &  2.0 & 443 \\
& Second peak &9-13&  -23.77675 &  231.10693 &  -36.28714 &  441.03471 &   90.70316 & 1.7 & 443 \\
&             &13-17&  -23.08951 &  212.20550 &  -35.21583 &  406.45133 &   92.79225 & 1.3 & 443 \\
&             &17-22&  -18.24024 &  122.91978 &  -28.28962 &  237.51385 &   69.65176 & 1.5 & 443 \\
&             &22-27&  -17.27004 &  105.47879 &  -26.53128 &  203.78858 &   62.65549 & 1.4 & 443 \\
&             &27-60&  -17.32346 &  102.82002 &  -25.53566 &  184.54082 &   63.96669 & 1.1 & 443 \\

\hline

\end{tabular}
\end{center}
\end{table*}

\begin{table*}

\footnotesize
%%\tiny
\caption {The phase-resolved spectral analysis results of the Crab pulsar for power law model}
\scriptsize{}\label{table:Crabphrespec}
\medskip
\begin{center}
\begin{tabular}{c c c c c c c}
\hline \hline

& Pulse phase range & Normalization & Spectral Index & $\chi^{2}$ & d.o.f. \\
\hline
& -0.140...-0.130 & 0.045 $\pm$ 0.006 & 1.721 $\pm$ 0.098 & 95.4 & 86 \\
& -0.130...-0.120 & 0.060 $\pm$ 0.005 & 1.708 $\pm$ 0.055 & 91.4 & 86 \\
& -0.120...-0.110 & 0.098 $\pm$ 0.007 & 1.735 $\pm$ 0.039 & 92.3 & 86 \\
& -0.110...-0.100 & 0.147 $\pm$ 0.007 & 1.750 $\pm$ 0.025 & 91.3 & 86 \\
& -0.100...-0.095 & 0.195 $\pm$ 0.011 & 1.784 $\pm$ 0.031 & 91.4 & 86 \\
& -0.095...-0.090 & 0.226 $\pm$ 0.011 & 1.777 $\pm$ 0.026 & 89.3 & 86 \\
& -0.090...-0.085 & 0.276 $\pm$ 0.012 & 1.784 $\pm$ 0.022 & 89.7 & 86 \\
& -0.085...-0.080 & 0.346 $\pm$ 0.013 & 1.813 $\pm$ 0.020 & 93.7 & 86 \\
& -0.080...-0.075 & 0.421 $\pm$ 0.013 & 1.830 $\pm$ 0.016 & 94.8 & 86 \\
& -0.075...-0.070 & 0.495 $\pm$ 0.014 & 1.829 $\pm$ 0.014 & 93.5 & 86 \\
& -0.070...-0.065 & 0.585 $\pm$ 0.014 & 1.836 $\pm$ 0.012 & 94.9 & 86 \\
& -0.065...-0.060 & 0.711 $\pm$ 0.016 & 1.856 $\pm$ 0.011 & 90.7 & 86 \\
& -0.060...-0.055 & 0.854 $\pm$ 0.019 & 1.870 $\pm$ 0.010 & 93.1 & 86 \\
& -0.055...-0.050 & 1.007 $\pm$ 0.018 & 1.876 $\pm$ 0.008 & 93.6 & 86 \\
& -0.050...-0.045 & 1.202 $\pm$ 0.021 & 1.888 $\pm$ 0.007 & 97.4 & 86 \\
& -0.045...-0.040 & 1.427 $\pm$ 0.022 & 1.900 $\pm$ 0.006 & 96.6 & 86 \\
& -0.040...-0.035 & 1.672 $\pm$ 0.024 & 1.903 $\pm$ 0.005 & 99.3 & 86 \\
& -0.035...-0.030 & 1.985 $\pm$ 0.027 & 1.910 $\pm$ 0.005 & 101.7 & 86 \\
& -0.030...-0.025 & 2.343 $\pm$ 0.028 & 1.914 $\pm$ 0.004 & 100.2 & 86 \\
& -0.025...-0.020 & 2.801 $\pm$ 0.032 & 1.921 $\pm$ 0.004 & 103.5 & 86 \\
& -0.020...-0.015 & 3.407 $\pm$ 0.039 & 1.933 $\pm$ 0.003 & 109.8 & 86 \\
& -0.015...-0.010 & 4.114 $\pm$ 0.045 & 1.941 $\pm$ 0.003 & 115.3 & 86 \\
& -0.010...-0.005 & 4.900 $\pm$ 0.049 & 1.948 $\pm$ 0.003 & 120.5 & 86 \\
& -0.005...-0.000 & 5.382 $\pm$ 0.049 & 1.949 $\pm$ 0.002 & 119.8 & 86 \\
& -0.000...0.005 & 5.108 $\pm$ 0.042 & 1.947 $\pm$ 0.003 & 120.5 & 86 \\
& 0.005...0.010 & 4.047 $\pm$ 0.044 & 1.933 $\pm$ 0.003 & 109.6 & 86 \\
& 0.010...0.015 & 2.958 $\pm$ 0.038 & 1.909 $\pm$ 0.004 & 103.2 & 86 \\
& 0.015...0.020 & 2.189 $\pm$ 0.032 & 1.882 $\pm$ 0.005 & 100.2 & 86 \\
& 0.020...0.025 & 1.693 $\pm$ 0.022 & 1.862 $\pm$ 0.005 & 97.4 & 86 \\
& 0.025...0.030 & 1.353 $\pm$ 0.019 & 1.844 $\pm$ 0.006 & 91.0 & 86 \\
& 0.030...0.035 & 1.081 $\pm$ 0.018 & 1.814 $\pm$ 0.007 & 92.0 & 86 \\
& 0.035...0.040 & 0.904 $\pm$ 0.016 & 1.800 $\pm$ 0.008 & 99.2 & 86 \\
& 0.040...0.045 & 0.772 $\pm$ 0.015 & 1.786 $\pm$ 0.009 & 92.9 & 86 \\
& 0.045...0.050 & 0.667 $\pm$ 0.014 & 1.768 $\pm$ 0.009 & 94.3 & 86 \\
& 0.050...0.055 & 0.592 $\pm$ 0.013 & 1.760 $\pm$ 0.011 & 90.4 & 86 \\
& 0.055...0.060 & 0.519 $\pm$ 0.012 & 1.739 $\pm$ 0.012 & 93.2 & 86 \\
& 0.060...0.065 & 0.471 $\pm$ 0.012 & 1.728 $\pm$ 0.012 & 94.4 & 86 \\
& 0.065...0.070 & 0.423 $\pm$ 0.011 & 1.718 $\pm$ 0.013 & 92.6 & 86 \\
& 0.070...0.075 & 0.382 $\pm$ 0.011 & 1.699 $\pm$ 0.013 & 95.7 & 86 \\
& 0.075...0.080 & 0.355 $\pm$ 0.010 & 1.694 $\pm$ 0.014 & 92.4 & 86 \\
& 0.080...0.090 & 0.305 $\pm$ 0.008 & 1.670 $\pm$ 0.011 & 92.6 & 86 \\
& 0.090...0.100 & 0.274 $\pm$ 0.008 & 1.661 $\pm$ 0.012 & 95.8 & 86 \\
& 0.100...0.110 & 0.250 $\pm$ 0.007 & 1.649 $\pm$ 0.013 & 92.5 & 86 \\
& 0.110...0.120 & 0.227 $\pm$ 0.007 & 1.623 $\pm$ 0.013 & 89.7 & 86 \\
& 0.120...0.130 & 0.208 $\pm$ 0.006 & 1.604 $\pm$ 0.013 & 92.9 & 86 \\
& 0.130...0.140 & 0.205 $\pm$ 0.006 & 1.608 $\pm$ 0.015 & 91.7 & 86 \\
& 0.140...0.150 & 0.198 $\pm$ 0.006 & 1.595 $\pm$ 0.014 & 93.6 & 86 \\
& 0.150...0.160 & 0.201 $\pm$ 0.006 & 1.602 $\pm$ 0.014 & 90.3 & 86 \\
& 0.160...0.170 & 0.197 $\pm$ 0.006 & 1.592 $\pm$ 0.014 & 90.4 & 86 \\
& 0.170...0.180 & 0.203 $\pm$ 0.006 & 1.592 $\pm$ 0.014 & 90.1 & 86 \\
& 0.180...0.190 & 0.212 $\pm$ 0.006 & 1.597 $\pm$ 0.013 & 89.6 & 86 \\
& 0.190...0.200 & 0.223 $\pm$ 0.006 & 1.601 $\pm$ 0.013 & 93.1 & 86 \\
& 0.200...0.210 & 0.232 $\pm$ 0.006 & 1.596 $\pm$ 0.012 & 91.7 & 86 \\
& 0.210...0.220 & 0.247 $\pm$ 0.006 & 1.597 $\pm$ 0.011 & 93.5 & 86 \\
& 0.220...0.230 & 0.270 $\pm$ 0.006 & 1.607 $\pm$ 0.011 & 95.3 & 86 \\
& 0.230...0.240 & 0.297 $\pm$ 0.007 & 1.611 $\pm$ 0.009 & 93.1 & 86 \\
\hline

\end{tabular}
\end{center}

\end{table*}

\begin{table*}

\footnotesize
%%\tiny
\caption {The phase-resolved spectral analysis results of the Crab
pulsar for power law model-continued} \scriptsize{}\label{table:Crabphrespec2}
\medskip
\begin{center}
\begin{tabular}{c c c c c c c}
\hline \hline

& Pulse phase range & Normalization & Spectral Index & $\chi^{2}$ &d.o.f.\\

\hline
& 0.240...0.250 & 0.321 $\pm$ 0.007 & 1.617 $\pm$ 0.009 & 91.0 & 86 \\
& 0.250...0.260 & 0.355 $\pm$ 0.007 & 1.623 $\pm$ 0.009 & 97.2 & 86 \\
& 0.260...0.270 & 0.396 $\pm$ 0.007 & 1.632 $\pm$ 0.008 & 93.8 & 86 \\
& 0.270...0.280 & 0.447 $\pm$ 0.008 & 1.643 $\pm$ 0.007 & 95.9 & 86 \\
& 0.280...0.290 & 0.496 $\pm$ 0.008 & 1.647 $\pm$ 0.007 & 96.1 & 86 \\
& 0.290...0.300 & 0.571 $\pm$ 0.009 & 1.662 $\pm$ 0.006 & 97.4 & 86 \\
& 0.300...0.305 & 0.664 $\pm$ 0.011 & 1.682 $\pm$ 0.008 & 95.3 & 86 \\
& 0.305...0.310 & 0.730 $\pm$ 0.013 & 1.698 $\pm$ 0.007 & 97.8 & 86 \\
& 0.310...0.315 & 0.765 $\pm$ 0.012 & 1.696 $\pm$ 0.007 & 93.7 & 86 \\
& 0.315...0.320 & 0.826 $\pm$ 0.013 & 1.704 $\pm$ 0.006 & 96.7 & 86 \\
& 0.320...0.325 & 0.905 $\pm$ 0.013 & 1.720 $\pm$ 0.006 & 97.2 & 86 \\
& 0.325...0.330 & 0.973 $\pm$ 0.014 & 1.725 $\pm$ 0.006 & 94.4 & 86 \\
& 0.330...0.335 & 1.046 $\pm$ 0.015 & 1.732 $\pm$ 0.006 & 99.1 & 86 \\
& 0.335...0.340 & 1.131 $\pm$ 0.015 & 1.739 $\pm$ 0.005 & 99.1 & 86 \\
& 0.340...0.345 & 1.225 $\pm$ 0.016 & 1.749 $\pm$ 0.005 & 101.0 & 86 \\
& 0.345...0.350 & 1.318 $\pm$ 0.017 & 1.752 $\pm$ 0.005 & 101.9 & 86 \\
& 0.350...0.355 & 1.422 $\pm$ 0.019 & 1.759 $\pm$ 0.005 & 99.8 & 86 \\
& 0.355...0.360 & 1.548 $\pm$ 0.017 & 1.766 $\pm$ 0.004 & 101.8 & 86 \\
& 0.360...0.365 & 1.671 $\pm$ 0.020 & 1.768 $\pm$ 0.004 & 98.0 & 86 \\
& 0.365...0.370 & 1.812 $\pm$ 0.021 & 1.776 $\pm$ 0.004 & 103.7 & 86 \\
& 0.370...0.375 & 1.962 $\pm$ 0.021 & 1.779 $\pm$ 0.003 & 103.2 & 86 \\
& 0.375...0.380 & 2.102 $\pm$ 0.022 & 1.779 $\pm$ 0.004 & 102.2 & 86 \\
& 0.380...0.385 & 2.265 $\pm$ 0.024 & 1.783 $\pm$ 0.003 & 100.6 & 86 \\
& 0.385...0.390 & 2.397 $\pm$ 0.024 & 1.784 $\pm$ 0.003 & 102.7 & 86 \\
& 0.390...0.395 & 2.465 $\pm$ 0.024 & 1.784 $\pm$ 0.003 & 103.7 & 86 \\
& 0.395...0.400 & 2.471 $\pm$ 0.024 & 1.789 $\pm$ 0.003 & 103.9 & 86 \\
& 0.400...0.405 & 2.344 $\pm$ 0.023 & 1.800 $\pm$ 0.003 & 106.7 & 86 \\
& 0.405...0.410 & 2.067 $\pm$ 0.022 & 1.812 $\pm$ 0.004 & 97.7 & 86 \\
& 0.410...0.415 & 1.778 $\pm$ 0.020 & 1.832 $\pm$ 0.004 & 98.1 & 86 \\
& 0.415...0.420 & 1.528 $\pm$ 0.021 & 1.857 $\pm$ 0.005 & 100.1 & 86 \\
& 0.420...0.425 & 1.298 $\pm$ 0.019 & 1.871 $\pm$ 0.006 & 95.3 & 86 \\
& 0.425...0.430 & 1.128 $\pm$ 0.018 & 1.889 $\pm$ 0.007 & 91.5 & 86 \\
& 0.430...0.435 & 0.980 $\pm$ 0.019 & 1.908 $\pm$ 0.009 & 93.4 & 86 \\
& 0.435...0.440 & 0.852 $\pm$ 0.018 & 1.916 $\pm$ 0.011 & 91.2 & 86 \\
& 0.440...0.445 & 0.766 $\pm$ 0.018 & 1.936 $\pm$ 0.011 & 91.9 & 86 \\
& 0.445...0.450 & 0.685 $\pm$ 0.018 & 1.959 $\pm$ 0.014 & 91.5 & 86 \\
& 0.450...0.455 & 0.589 $\pm$ 0.017 & 1.956 $\pm$ 0.016 & 90.5 & 86 \\
& 0.455...0.460 & 0.531 $\pm$ 0.017 & 1.976 $\pm$ 0.017 & 90.8 & 86 \\
& 0.460...0.465 & 0.462 $\pm$ 0.018 & 1.963 $\pm$ 0.020 & 91.7 & 86 \\
& 0.465...0.470 & 0.431 $\pm$ 0.019 & 1.984 $\pm$ 0.024 & 92.6 & 86 \\
& 0.470...0.475 & 0.378 $\pm$ 0.017 & 2.002 $\pm$ 0.026 & 93.6 & 86 \\
& 0.475...0.480 & 0.361 $\pm$ 0.018 & 2.028 $\pm$ 0.031 & 92.6 & 86 \\
& 0.480...0.485 & 0.313 $\pm$ 0.018 & 2.009 $\pm$ 0.032 & 91.4 & 86 \\
& 0.485...0.490 & 0.300 $\pm$ 0.018 & 2.050 $\pm$ 0.038 & 87.7 & 86 \\
& 0.490...0.495 & 0.271 $\pm$ 0.017 & 2.068 $\pm$ 0.040 & 87.6 & 86 \\
& 0.495...0.500 & 0.255 $\pm$ 0.019 & 2.084 $\pm$ 0.051 & 96.1 & 86 \\
& 0.500...0.510 & 0.209 $\pm$ 0.013 & 2.056 $\pm$ 0.039 & 93.4 & 86 \\
& 0.510...0.520 & 0.187 $\pm$ 0.013 & 2.120 $\pm$ 0.048 & 90.6 & 86 \\
& 0.520...0.530 & 0.156 $\pm$ 0.013 & 2.127 $\pm$ 0.058 & 91.8 & 86 \\
& 0.530...0.540 & 0.132 $\pm$ 0.015 & 2.117 $\pm$ 0.081 & 89.2 & 86 \\
& 0.540...0.550 & 0.120 $\pm$ 0.017 & 2.208 $\pm$ 0.102 & 93.4 & 86 \\

\hline

\end{tabular}
\end{center}
\end{table*}

\begin{table*}
\footnotesize
%%\tiny
\caption {The slopes of the spectral indices of the Crab
pulsar change with phase
} \scriptsize{}\label{table:Crabphaseslope}
\medskip
\begin{center}
\begin{tabular}{c c c}
\hline \hline
&phase range& slope\\
\hline
& -0.14 - -0.04 & 2.17 \\
& -0.04 - -0.01 & 1.55 \\
& -0.01 - 0.00 & -0.17 \\
& 0.00 - 0.04 & -4.36 \\
& 0.04 - 0.14 & -2.05 \\
& 0.14 - 0.24 & 0.13 \\
& 0.24 - 0.36 & 1.43 \\
& 0.36 - 0.40 & 0.31 \\
& 0.40 - 0.45 & 3.43 \\
& 0.45 - 0.60 & 2.44 \\
\hline
\end{tabular}
\end{center}
\end{table*}

\begin{table*}

\footnotesize
%%\tiny
\caption {The phase-resolved spectral analysis results of the Crab
pulsar for log-parabola model} \scriptsize{}\label{table:Crabphrespeclogpar}
\medskip
\begin{center}
\begin{tabular}{c c c c c c c c c}
\hline \hline

& Pulse phase range & Normalization & $\alpha$ &   $\beta$   & $\chi^{2}$ &d.o.f. & $\Delta\chi^{2}$ & Propability\\

\hline
& -0.140...-0.130 & 0.075 $\pm$ 0.012 & 1.251 $\pm$ 0.193 & 0.271 $\pm$ 0.193 &  94.7 & 85 & 0.8 & 4.2e-1 \\
& -0.130...-0.120 & 0.079 $\pm$ 0.012 & 1.134 $\pm$ 0.165 & 0.322 $\pm$ 0.158 &  90.3 & 85 & 1.1 & 3.2e-1 \\
& -0.120...-0.110 & 0.056 $\pm$ 0.008 & 0.826 $\pm$ 0.122 & 0.492 $\pm$ 0.130 &  90.1 & 85 & 2.1 & 1.5e-1 \\
& -0.110...-0.100 & 0.081 $\pm$ 0.011 & 0.971 $\pm$ 0.131 & 0.412 $\pm$ 0.097 &  88.5 & 85 & 2.9 & 1.1e-1 \\
& -0.100...-0.095 & 0.113 $\pm$ 0.016 & 1.046 $\pm$ 0.140 & 0.400 $\pm$ 0.110 &  89.3 & 85 & 2.1 & 1.6e-1 \\
& -0.095...-0.090 & 0.133 $\pm$ 0.018 & 1.120 $\pm$ 0.149 & 0.361 $\pm$ 0.094 &  87.5 & 85 & 1.8 & 1.9e-1 \\
& -0.090...-0.085 & 0.192 $\pm$ 0.022 & 1.296 $\pm$ 0.149 & 0.255 $\pm$ 0.078 &  88.0 & 85 & 1.8 & 2.0e-1 \\
& -0.085...-0.080 & 0.212 $\pm$ 0.021 & 1.237 $\pm$ 0.137 & 0.298 $\pm$ 0.068 &  91.4 & 85 & 2.3 & 1.5e-1 \\
& -0.080...-0.075 & 0.223 $\pm$ 0.020 & 1.136 $\pm$ 0.119 & 0.363 $\pm$ 0.060 &  91.2 & 85 & 3.6 & 7.0e-2 \\
& -0.075...-0.070 & 0.297 $\pm$ 0.024 & 1.258 $\pm$ 0.112 & 0.296 $\pm$ 0.054 &  89.7 & 85 & 3.7 & 6.1e-2 \\
& -0.070...-0.065 & 0.377 $\pm$ 0.028 & 1.357 $\pm$ 0.098 & 0.249 $\pm$ 0.045 &  91.3 & 85 & 3.7 & 6.9e-2 \\
& -0.065...-0.060 & 0.511 $\pm$ 0.033 & 1.498 $\pm$ 0.084 & 0.187 $\pm$ 0.040 &  87.7 & 85 & 3.0 & 9.0e-2 \\
& -0.060...-0.055 & 0.620 $\pm$ 0.038 & 1.511 $\pm$ 0.075 & 0.187 $\pm$ 0.037 &  89.3 & 85 & 3.8 & 6.0e-2 \\
& -0.055...-0.050 & 0.662 $\pm$ 0.034 & 1.452 $\pm$ 0.061 & 0.222 $\pm$ 0.029 &  88.1 & 85 & 5.6 & 2.3e-2 \\
& -0.050...-0.045 & 0.860 $\pm$ 0.044 & 1.537 $\pm$ 0.056 & 0.184 $\pm$ 0.027 &  91.8 & 85 & 5.6 & 2.5e-2 \\
& -0.045...-0.040 & 1.032 $\pm$ 0.042 & 1.572 $\pm$ 0.046 & 0.172 $\pm$ 0.023 &  90.6 & 85 & 5.9 & 2.0e-2 \\
& -0.040...-0.035 & 1.223 $\pm$ 0.044 & 1.590 $\pm$ 0.039 & 0.164 $\pm$ 0.020 &  92.8 & 85 & 6.4 & 1.7e-2 \\
& -0.035...-0.030 & 1.431 $\pm$ 0.045 & 1.585 $\pm$ 0.034 & 0.171 $\pm$ 0.017 &  92.5 & 85 & 9.2 & 4.6e-3 \\
& -0.030...-0.025 & 1.643 $\pm$ 0.048 & 1.561 $\pm$ 0.031 & 0.186 $\pm$ 0.016 &  86.2 & 85 & 14.0 & 3.7e-4 \\
& -0.025...-0.020 & 2.018 $\pm$ 0.054 & 1.594 $\pm$ 0.027 & 0.172 $\pm$ 0.013 &  88.2 & 85 & 15.3 & 2.4e-4 \\
& -0.020...-0.015 & 2.464 $\pm$ 0.061 & 1.609 $\pm$ 0.023 & 0.171 $\pm$ 0.012 &  90.2 & 85 & 19.6 & 4.5e-5 \\
& -0.015...-0.010 & 2.999 $\pm$ 0.064 & 1.627 $\pm$ 0.020 & 0.166 $\pm$ 0.010 &  90.8 & 85 & 24.5 & 7.1e-6 \\
& -0.010...-0.005 & 3.647 $\pm$ 0.071 & 1.655 $\pm$ 0.017 & 0.155 $\pm$ 0.009 &  94.0 & 85 & 26.3 & 4.7e-6 \\
& -0.005...-0.000 & 4.038 $\pm$ 0.073 & 1.664 $\pm$ 0.016 & 0.150 $\pm$ 0.008 &  91.4 & 85 & 28.4 & 1.7e-6 \\
& -0.000...0.005 & 3.862 $\pm$ 0.067 & 1.670 $\pm$ 0.017 & 0.146 $\pm$ 0.009 &  94.7 & 85 & 25.8 & 6.3e-6 \\
& 0.005...0.010 & 3.070 $\pm$ 0.064 & 1.659 $\pm$ 0.020 & 0.144 $\pm$ 0.010 &  90.7 & 85 & 18.9 & 6.3e-5 \\
& 0.010...0.015 & 2.212 $\pm$ 0.053 & 1.623 $\pm$ 0.024 & 0.150 $\pm$ 0.012 &  89.7 & 85 & 13.4 & 5.7e-4 \\
& 0.015...0.020 & 1.676 $\pm$ 0.052 & 1.615 $\pm$ 0.031 & 0.140 $\pm$ 0.015 &  91.7 & 85 & 8.6 & 6.1e-3 \\
& 0.020...0.025 & 1.315 $\pm$ 0.047 & 1.604 $\pm$ 0.038 & 0.134 $\pm$ 0.019 &  91.5 & 85 & 5.9 & 2.1e-2 \\
& 0.025...0.030 & 1.040 $\pm$ 0.041 & 1.580 $\pm$ 0.043 & 0.137 $\pm$ 0.021 &  86.2 & 85 & 4.9 & 3.2e-2 \\
& 0.030...0.035 & 0.813 $\pm$ 0.036 & 1.528 $\pm$ 0.050 & 0.148 $\pm$ 0.025 &  87.5 & 85 & 4.4 & 3.9e-2 \\
& 0.035...0.040 & 0.665 $\pm$ 0.033 & 1.485 $\pm$ 0.057 & 0.162 $\pm$ 0.027 &  95.2 & 85 & 4.0 & 6.1e-2 \\
& 0.040...0.045 & 0.575 $\pm$ 0.034 & 1.469 $\pm$ 0.069 & 0.163 $\pm$ 0.032 &  89.0 & 85 & 3.8 & 5.7e-2 \\
& 0.045...0.050 & 0.468 $\pm$ 0.028 & 1.395 $\pm$ 0.075 & 0.191 $\pm$ 0.035 &  90.1 & 85 & 4.2 & 4.9e-2 \\
& 0.050...0.055 & 0.433 $\pm$ 0.028 & 1.421 $\pm$ 0.080 & 0.173 $\pm$ 0.037 &  87.5 & 85 & 2.8 & 9.8e-2 \\
& 0.055...0.060 & 0.388 $\pm$ 0.026 & 1.407 $\pm$ 0.085 & 0.169 $\pm$ 0.040 &  90.6 & 85 & 2.6 & 1.2e-1 \\
& 0.060...0.065 & 0.311 $\pm$ 0.022 & 1.308 $\pm$ 0.087 & 0.213 $\pm$ 0.040 &  91.6 & 85 & 2.8 & 1.1e-1 \\
& 0.065...0.070 & 0.308 $\pm$ 0.023 & 1.364 $\pm$ 0.098 & 0.178 $\pm$ 0.045 &  90.4 & 85 & 2.2 & 1.6e-1 \\
& 0.070...0.075 & 0.255 $\pm$ 0.021 & 1.230 $\pm$ 0.107 & 0.237 $\pm$ 0.049 &  92.4 & 85 & 3.3 & 8.7e-2 \\
& 0.075...0.080 & 0.250 $\pm$ 0.022 & 1.260 $\pm$ 0.118 & 0.221 $\pm$ 0.055 &  89.3 & 85 & 3.1 & 9.1e-2 \\
& 0.080...0.090 & 0.208 $\pm$ 0.014 & 1.263 $\pm$ 0.086 & 0.204 $\pm$ 0.040 &  89.4 & 85 & 3.2 & 8.3e-2 \\
& 0.090...0.100 & 0.209 $\pm$ 0.015 & 1.360 $\pm$ 0.093 & 0.151 $\pm$ 0.042 &  93.6 & 85 & 2.2 & 1.6e-1 \\
& 0.100...0.110 & 0.168 $\pm$ 0.013 & 1.235 $\pm$ 0.096 & 0.209 $\pm$ 0.044 &  89.8 & 85 & 2.7 & 1.2e-1 \\
& 0.110...0.120 & 0.148 $\pm$ 0.012 & 1.158 $\pm$ 0.107 & 0.232 $\pm$ 0.047 &  86.4 & 85 & 3.4 & 7.5e-2 \\
& 0.120...0.130 & 0.137 $\pm$ 0.011 & 1.162 $\pm$ 0.102 & 0.219 $\pm$ 0.049 &  90.0 & 85 & 3.0 & 1.0e-1 \\
& 0.130...0.140 & 0.145 $\pm$ 0.012 & 1.232 $\pm$ 0.102 & 0.186 $\pm$ 0.047 &  89.7 & 85 & 2.1 & 1.7e-1 \\
& 0.140...0.150 & 0.148 $\pm$ 0.012 & 1.242 $\pm$ 0.108 & 0.175 $\pm$ 0.050 &  91.3 & 85 & 2.3 & 1.5e-1 \\
& 0.150...0.160 & 0.139 $\pm$ 0.011 & 1.177 $\pm$ 0.103 & 0.211 $\pm$ 0.051 &  87.7 & 85 & 2.6 & 1.1e-1 \\
& 0.160...0.170 & 0.141 $\pm$ 0.011 & 1.222 $\pm$ 0.102 & 0.183 $\pm$ 0.048 &  88.3 & 85 & 2.0 & 1.6e-1 \\
& 0.170...0.180 & 0.135 $\pm$ 0.011 & 1.143 $\pm$ 0.106 & 0.222 $\pm$ 0.048 &  87.2 & 85 & 2.9 & 9.4e-2 \\
& 0.180...0.190 & 0.132 $\pm$ 0.010 & 1.083 $\pm$ 0.099 & 0.253 $\pm$ 0.049 &  85.9 & 85 & 3.6 & 6.1e-2 \\
& 0.190...0.200 & 0.153 $\pm$ 0.012 & 1.194 $\pm$ 0.097 & 0.201 $\pm$ 0.044 &  90.4 & 85 & 2.8 & 1.1e-1 \\
& 0.200...0.210 & 0.154 $\pm$ 0.011 & 1.166 $\pm$ 0.093 & 0.213 $\pm$ 0.043 &  88.4 & 85 & 3.3 & 7.7e-2 \\
& 0.210...0.220 & 0.175 $\pm$ 0.012 & 1.225 $\pm$ 0.090 & 0.184 $\pm$ 0.040 &  90.3 & 85 & 3.1 & 8.9e-2 \\
& 0.220...0.230 & 0.171 $\pm$ 0.011 & 1.161 $\pm$ 0.078 & 0.221 $\pm$ 0.036 &  91.7 & 85 & 3.6 & 7.0e-2 \\
& 0.230...0.240 & 0.191 $\pm$ 0.012 & 1.167 $\pm$ 0.077 & 0.220 $\pm$ 0.035 &  88.7 & 85 & 4.3 & 4.3e-2 \\

\hline

\end{tabular}
\end{center}

Note: $\Delta\chi^{2}$ is the reduced value of the $\chi^{2}$ when using the log-parabola model instead of power-law model

\end{table*}

\begin{table*}

\footnotesize
%%\tiny
\caption {The phase-resolved spectral analysis results of the Crab
pulsar for log-parabola model-continued} \scriptsize{}\label{table:Crabphrespeclogpar2}
\medskip
\begin{center}
\begin{tabular}{c c c c c c c c c}
\hline \hline

& Pulse phase range & Normalization & $\alpha$ &   $\beta$   & $\chi^{2}$ &d.o.f. & $\Delta\chi^{2}$ & propability\\

\hline

& 0.240...0.250 & 0.207 $\pm$ 0.012 & 1.188 $\pm$ 0.068 & 0.212 $\pm$ 0.031 &  86.8 & 85 & 4.2 & 4.5e-2 \\
& 0.250...0.260 & 0.230 $\pm$ 0.014 & 1.185 $\pm$ 0.069 & 0.218 $\pm$ 0.031 &  91.4 & 85 & 5.9 & 2.2e-2 \\
& 0.260...0.270 & 0.276 $\pm$ 0.014 & 1.282 $\pm$ 0.058 & 0.175 $\pm$ 0.028 &  89.6 & 85 & 4.1 & 5.0e-2 \\
& 0.270...0.280 & 0.305 $\pm$ 0.015 & 1.264 $\pm$ 0.057 & 0.189 $\pm$ 0.027 &  90.0 & 85 & 6.0 & 2.0e-2 \\
& 0.280...0.290 & 0.344 $\pm$ 0.015 & 1.300 $\pm$ 0.046 & 0.173 $\pm$ 0.022 &  90.8 & 85 & 5.3 & 2.9e-2 \\
& 0.290...0.300 & 0.390 $\pm$ 0.017 & 1.292 $\pm$ 0.046 & 0.185 $\pm$ 0.022 &  89.6 & 85 & 7.8 & 8.0e-3 \\
& 0.300...0.305 & 0.453 $\pm$ 0.023 & 1.306 $\pm$ 0.058 & 0.190 $\pm$ 0.027 &  90.3 & 85 & 5.0 & 3.3e-2 \\
& 0.305...0.310 & 0.466 $\pm$ 0.023 & 1.256 $\pm$ 0.056 & 0.223 $\pm$ 0.027 &  90.6 & 85 & 7.2 & 1.1e-2 \\
& 0.310...0.315 & 0.536 $\pm$ 0.024 & 1.352 $\pm$ 0.050 & 0.173 $\pm$ 0.024 &  88.7 & 85 & 5.0 & 3.1e-2 \\
& 0.315...0.320 & 0.559 $\pm$ 0.026 & 1.320 $\pm$ 0.052 & 0.194 $\pm$ 0.025 &  89.7 & 85 & 6.9 & 1.2e-2 \\
& 0.320...0.325 & 0.600 $\pm$ 0.025 & 1.322 $\pm$ 0.046 & 0.202 $\pm$ 0.022 &  89.7 & 85 & 7.5 & 9.0e-3 \\
& 0.325...0.330 & 0.711 $\pm$ 0.029 & 1.415 $\pm$ 0.046 & 0.157 $\pm$ 0.022 &  88.5 & 85 & 5.9 & 1.9e-2 \\
& 0.330...0.335 & 0.731 $\pm$ 0.029 & 1.380 $\pm$ 0.043 & 0.179 $\pm$ 0.021 &  91.6 & 85 & 7.5 & 9.7e-3 \\
& 0.335...0.340 & 0.781 $\pm$ 0.028 & 1.382 $\pm$ 0.039 & 0.182 $\pm$ 0.019 &  90.5 & 85 & 8.6 & 5.6e-3 \\
& 0.340...0.345 & 0.850 $\pm$ 0.032 & 1.390 $\pm$ 0.040 & 0.183 $\pm$ 0.019 &  90.9 & 85 & 10.0 & 2.9e-3 \\
& 0.345...0.350 & 0.946 $\pm$ 0.034 & 1.424 $\pm$ 0.039 & 0.168 $\pm$ 0.019 &  92.4 & 85 & 9.6 & 4.1e-3 \\
& 0.350...0.355 & 1.013 $\pm$ 0.033 & 1.429 $\pm$ 0.035 & 0.168 $\pm$ 0.017 &  89.7 & 85 & 10.1 & 2.7e-3 \\
& 0.355...0.360 & 1.127 $\pm$ 0.037 & 1.452 $\pm$ 0.034 & 0.160 $\pm$ 0.017 &  91.1 & 85 & 10.6 & 2.2e-3 \\
& 0.360...0.365 & 1.231 $\pm$ 0.037 & 1.469 $\pm$ 0.030 & 0.153 $\pm$ 0.015 &  87.9 & 85 & 10.1 & 2.4e-3 \\
& 0.365...0.370 & 1.309 $\pm$ 0.039 & 1.457 $\pm$ 0.030 & 0.163 $\pm$ 0.015 &  90.6 & 85 & 13.1 & 7.2e-4 \\
& 0.370...0.375 & 1.420 $\pm$ 0.037 & 1.466 $\pm$ 0.027 & 0.161 $\pm$ 0.014 &  88.7 & 85 & 14.4 & 3.5e-4 \\
& 0.375...0.380 & 1.581 $\pm$ 0.042 & 1.500 $\pm$ 0.026 & 0.143 $\pm$ 0.013 &  89.6 & 85 & 12.5 & 8.7e-4 \\
& 0.380...0.385 & 1.712 $\pm$ 0.042 & 1.510 $\pm$ 0.023 & 0.140 $\pm$ 0.012 &  87.6 & 85 & 12.9 & 6.3e-4 \\
& 0.385...0.390 & 1.836 $\pm$ 0.043 & 1.524 $\pm$ 0.023 & 0.133 $\pm$ 0.011 &  89.6 & 85 & 13.0 & 6.9e-4 \\
& 0.390...0.395 & 1.875 $\pm$ 0.044 & 1.517 $\pm$ 0.023 & 0.137 $\pm$ 0.012 &  89.3 & 85 & 14.4 & 3.8e-4 \\
& 0.395...0.400 & 1.886 $\pm$ 0.043 & 1.527 $\pm$ 0.022 & 0.135 $\pm$ 0.011 &  90.2 & 85 & 13.6 & 5.6e-4 \\
& 0.400...0.405 & 1.748 $\pm$ 0.044 & 1.513 $\pm$ 0.024 & 0.148 $\pm$ 0.012 &  92.4 & 85 & 14.4 & 4.8e-4 \\
& 0.405...0.410 & 1.561 $\pm$ 0.043 & 1.536 $\pm$ 0.028 & 0.142 $\pm$ 0.014 &  86.7 & 85 & 10.9 & 1.5e-3 \\
& 0.410...0.415 & 1.343 $\pm$ 0.040 & 1.557 $\pm$ 0.032 & 0.143 $\pm$ 0.016 &  90.5 & 85 & 7.7 & 8.8e-3 \\
& 0.415...0.420 & 1.145 $\pm$ 0.047 & 1.560 $\pm$ 0.041 & 0.154 $\pm$ 0.020 &  93.5 & 85 & 6.6 & 1.6e-2 \\
& 0.420...0.425 & 0.908 $\pm$ 0.038 & 1.514 $\pm$ 0.047 & 0.186 $\pm$ 0.023 &  89.0 & 85 & 6.3 & 1.6e-2 \\
& 0.425...0.430 & 0.817 $\pm$ 0.038 & 1.563 $\pm$ 0.054 & 0.170 $\pm$ 0.027 &  87.5 & 85 & 4.0 & 5.2e-2 \\
& 0.430...0.435 & 0.688 $\pm$ 0.038 & 1.542 $\pm$ 0.066 & 0.192 $\pm$ 0.032 &  89.7 & 85 & 3.6 & 6.6e-2 \\
& 0.435...0.440 & 0.627 $\pm$ 0.038 & 1.592 $\pm$ 0.076 & 0.170 $\pm$ 0.037 &  88.8 & 85 & 2.4 & 1.3e-1 \\
& 0.440...0.445 & 0.553 $\pm$ 0.037 & 1.586 $\pm$ 0.087 & 0.184 $\pm$ 0.042 &  89.7 & 85 & 2.1 & 1.5e-1 \\
& 0.445...0.450 & 0.499 $\pm$ 0.038 & 1.597 $\pm$ 0.106 & 0.192 $\pm$ 0.051 &  89.5 & 85 & 2.0 & 1.7e-1 \\
& 0.450...0.455 & 0.404 $\pm$ 0.034 & 1.527 $\pm$ 0.119 & 0.227 $\pm$ 0.056 &  88.6 & 85 & 1.9 & 1.8e-1 \\
& 0.455...0.460 & 0.333 $\pm$ 0.031 & 1.434 $\pm$ 0.140 & 0.289 $\pm$ 0.067 &  88.5 & 85 & 2.2 & 1.4e-1 \\
& 0.460...0.465 & 0.322 $\pm$ 0.033 & 1.486 $\pm$ 0.153 & 0.259 $\pm$ 0.075 &  89.9 & 85 & 1.8 & 1.9e-1 \\
& 0.465...0.470 & 0.314 $\pm$ 0.038 & 1.528 $\pm$ 0.168 & 0.251 $\pm$ 0.083 &  91.2 & 85 & 1.4 & 2.5e-1 \\
& 0.470...0.475 & 0.258 $\pm$ 0.036 & 1.455 $\pm$ 0.176 & 0.301 $\pm$ 0.097 &  92.2 & 85 & 1.3 & 2.7e-1 \\
& 0.475...0.480 & 0.267 $\pm$ 0.041 & 1.493 $\pm$ 0.190 & 0.300 $\pm$ 0.107 &  91.2 & 85 & 1.3 & 2.5e-1 \\
& 0.480...0.485 & 0.198 $\pm$ 0.029 & 1.196 $\pm$ 0.160 & 0.454 $\pm$ 0.124 &  89.3 & 85 & 2.1 & 1.6e-1 \\
& 0.485...0.490 & 0.231 $\pm$ 0.035 & 1.475 $\pm$ 0.207 & 0.317 $\pm$ 0.126 &  86.7 & 85 & 1.0 & 3.1e-1 \\
& 0.490...0.495 & 0.226 $\pm$ 0.034 & 1.466 $\pm$ 0.211 & 0.328 $\pm$ 0.143 &  86.5 & 85 & 1.1 & 3.1e-1 \\
& 0.495...0.500 & 0.200 $\pm$ 0.030 & 1.312 $\pm$ 0.186 & 0.451 $\pm$ 0.161 &  94.9 & 85 & 1.3 & 2.9e-1 \\
& 0.500...0.510 & 0.252 $\pm$ 0.038 & 1.578 $\pm$ 0.203 & 0.268 $\pm$ 0.134 &  91.7 & 85 & 1.6 & 2.1e-1 \\
& 0.510...0.520 & 0.144 $\pm$ 0.022 & 1.262 $\pm$ 0.189 & 0.486 $\pm$ 0.160 &  89.2 & 85 & 1.4 & 2.5e-1 \\
& 0.520...0.530 & 0.109 $\pm$ 0.016 & 1.052 $\pm$ 0.163 & 0.640 $\pm$ 0.190 &  90.4 & 85 & 1.4 & 2.5e-1 \\
& 0.530...0.540 & 0.143 $\pm$ 0.022 & 1.335 $\pm$ 0.200 & 0.442 $\pm$ 0.195 &  88.3 & 85 & 0.9 & 3.5e-1 \\
& 0.540...0.550 & 0.122 $\pm$ 0.018 & 1.276 $\pm$ 0.187 & 0.576 $\pm$ 0.256 &  92.5 & 85 & 0.9 & 3.8e-1 \\
\hline
\end{tabular}
\end{center}

Note: $\Delta\chi^{2}$ is the reduced value of the $\chi^{2}$ when using the log-parabola model instead of power-law model

\end{table*}

\begin{table*}

\footnotesize
%%\tiny
\caption {The coefficients of two Gaussian functions of PSR B1509-58 profiles}
\scriptsize{}\label{table:B1509profit}
\medskip
\begin{center}
\begin{tabular}{c c c c c c c c c c}
\hline \hline
&&2-5\,keV& 5-9\,keV & 9-13\,keV & 13-18\,keV & 18-22\,keV & 22-27\,keV & $>27$\,keV  \\
\hline
&$N_{1}$ & 1.632 $\pm$ 0.097  & 1.098 $\pm$ 0.040  & 1.088 $\pm$ 0.048  & 1.554 $\pm$ 0.092  & 1.496 $\pm$ 0.119  & 1.509 $\pm$ 0.167  & 1.380 $\pm$ 0.222  \\
&$\sigma_{1}$& 0.055 $\pm$ 0.003  & 0.046 $\pm$ 0.002  & 0.046 $\pm$ 0.002  & 0.052 $\pm$ 0.003  & 0.050 $\pm$ 0.004  & 0.050 $\pm$ 0.005  & 0.042 $\pm$ 0.008  \\
&$\mu_{1}$& 0.246 $\pm$ 0.001  & 0.238 $\pm$ 0.001  & 0.239 $\pm$ 0.001  & 0.249 $\pm$ 0.001  & 0.251 $\pm$ 0.001  & 0.253 $\pm$ 0.002  & 0.259 $\pm$ 0.003  \\
&$N_{2}$ & 2.355 $\pm$ 0.049  & 2.500 $\pm$ 0.028  & 2.512 $\pm$ 0.033  & 2.494 $\pm$ 0.050  & 2.502 $\pm$ 0.068  & 2.661 $\pm$ 0.104  & 2.653 $\pm$ 0.151  \\
&$\sigma_{2}$ & 0.131 $\pm$ 0.003  & 0.136 $\pm$ 0.001  & 0.134 $\pm$ 0.002  & 0.126 $\pm$ 0.003  & 0.126 $\pm$ 0.004  & 0.127 $\pm$ 0.005  & 0.124 $\pm$ 0.006  \\
&$\mu_{2}$ & 0.381 $\pm$ 0.003  & 0.348 $\pm$ 0.001  & 0.348 $\pm$ 0.001  & 0.373 $\pm$ 0.002  & 0.371 $\pm$ 0.003  & 0.367 $\pm$ 0.004  & 0.359 $\pm$ 0.005  \\
&$\chi^{2}_{dof}$ &   2.2  &   2.1  &   2.1  &   2.3  &   1.6  &   1.1  &   1.0 \\
&$d.o.f$& 94  & 194  & 194  & 94  & 94  & 94  & 94 \\

\hline
\end{tabular}
\end{center}
\end{table*}

\begin{table*}
\footnotesize
%%\tiny
\caption {The phase-resolved spectral analysis results of PSR
B1509-58} \scriptsize{}\label{table:B1509phrespec}
\medskip
\begin{center}
\begin{tabular}{c c c c c c c}
\hline \hline

& Pulse phase Range & Normalization($10^{-3}$) & Spectral Index & $\chi^{2}_{dof}$ & d.o.f.\\
\hline
& 0.00-0.14 &  0.60$\pm$0.07 & 1.413$\pm$0.056 & 0.8 & 42 \\
& 0.14-0.16 &  2.95$\pm$0.31 & 1.453$\pm$0.043 & 0.9 & 96 \\
& 0.16-0.18 &  4.09$\pm$0.30 & 1.418$\pm$0.029 & 1.1 & 96 \\
& 0.18-0.20 &  5.85$\pm$0.31 & 1.419$\pm$0.023 & 0.9 & 96 \\
& 0.20-0.22 &  7.76$\pm$0.32 & 1.407$\pm$0.018 & 0.8 & 96 \\
& 0.22-0.24 &  9.61$\pm$0.34 & 1.393$\pm$0.015 & 0.9 & 96 \\
& 0.24-0.26 & 10.84$\pm$0.34 & 1.376$\pm$0.013 & 1.2 & 96 \\
& 0.26-0.28 & 10.78$\pm$0.32 & 1.342$\pm$0.013 & 1.2 & 96 \\
& 0.28-0.30 & 10.70$\pm$0.32 & 1.340$\pm$0.013 & 1.0 & 96 \\
& 0.30-0.32 & 10.29$\pm$0.32 & 1.344$\pm$0.013 & 1.1 & 96 \\
& 0.32-0.34 &  9.62$\pm$0.31 & 1.335$\pm$0.013 & 1.0 & 96 \\
& 0.34-0.36 &  8.99$\pm$0.30 & 1.328$\pm$0.014 & 1.2 & 96 \\
& 0.36-0.38 &  8.98$\pm$0.30 & 1.338$\pm$0.014 & 1.3 & 96 \\
& 0.38-0.40 &  8.88$\pm$0.30 & 1.348$\pm$0.015 & 1.1 & 96 \\
& 0.40-0.42 &  8.34$\pm$0.29 & 1.333$\pm$0.015 & 1.0 & 96 \\
& 0.42-0.44 &  8.14$\pm$0.30 & 1.351$\pm$0.016 & 1.2 & 96 \\
& 0.44-0.46 &  7.63$\pm$0.30 & 1.357$\pm$0.017 & 0.9 & 96 \\
& 0.46-0.48 &  6.73$\pm$0.28 & 1.352$\pm$0.018 & 1.1 & 96 \\
& 0.48-0.50 &  6.13$\pm$0.29 & 1.366$\pm$0.020 & 1.3 & 96 \\
& 0.50-0.52 &  5.48$\pm$0.29 & 1.381$\pm$0.023 & 1.1 & 96 \\
& 0.52-0.54 &  5.05$\pm$0.31 & 1.414$\pm$0.024 & 0.9 & 96 \\
& 0.54-0.56 &  4.25$\pm$0.30 & 1.411$\pm$0.028 & 0.9 & 96 \\
& 0.56-0.58 &  3.49$\pm$0.29 & 1.413$\pm$0.033 & 1.1 & 96 \\
& 0.58-0.60 &  3.08$\pm$0.30 & 1.433$\pm$0.039 & 1.2 & 96 \\
& 0.60-0.64 &  1.97$\pm$0.19 & 1.470$\pm$0.038 & 1.5 & 49 \\
& 0.64-0.80 &  0.73$\pm$0.08 & 1.460$\pm$0.049 & 1.6 & 40 \\
\hline

\end{tabular}
\end{center}

\end{table*}

\begin{table*}
\footnotesize
%%\tiny
\caption {The coefficients of two Gaussian functions of PSR B0540-69 profiles}
\scriptsize{}\label{table:B0540profit}
\medskip
\begin{center}
\begin{tabular}{c c c c c c c c c c}
\hline \hline
& & 2-4\,keV & 4-6\,keV & 6-8\,keV & 8-12\,keV & 12-16\,keV & 16-20\,keV & $>20$\,keV  \\
\hline
&$N_{1}$& 2.286 $\pm$ 0.136  & 2.269 $\pm$ 0.101  & 2.288 $\pm$ 0.102  & 2.288 $\pm$ 0.109  & 2.343 $\pm$ 0.147  & 2.214 $\pm$ 0.272  & 2.206 $\pm$ 1.104  \\
&$\sigma_{1}$ & 0.107 $\pm$ 0.011  & 0.105 $\pm$ 0.008  & 0.105 $\pm$ 0.008  & 0.104 $\pm$ 0.008  & 0.106 $\pm$ 0.012  & 0.110 $\pm$ 0.021  & 0.058 $\pm$ 0.034  \\
&$\mu_{1}$ & 0.323 $\pm$ 0.017  & 0.321 $\pm$ 0.013  & 0.321 $\pm$ 0.013  & 0.323 $\pm$ 0.013  & 0.326 $\pm$ 0.018  & 0.329 $\pm$ 0.035  & 0.297 $\pm$ 0.042  \\
&$N_{2}$& 1.755 $\pm$ 0.253  & 1.846 $\pm$ 0.187  & 1.806 $\pm$ 0.191  & 1.812 $\pm$ 0.200  & 1.801 $\pm$ 0.293  & 1.714 $\pm$ 0.574  & 2.719 $\pm$ 0.613  \\
&$\sigma_{2}$& 0.084 $\pm$ 0.010  & 0.083 $\pm$ 0.007  & 0.082 $\pm$ 0.007  & 0.083 $\pm$ 0.007  & 0.081 $\pm$ 0.009  & 0.081 $\pm$ 0.016  & 0.088 $\pm$ 0.045  \\
&$\mu_{2}$ & 0.532 $\pm$ 0.016  & 0.528 $\pm$ 0.011  & 0.527 $\pm$ 0.011  & 0.526 $\pm$ 0.012  & 0.526 $\pm$ 0.016  & 0.523 $\pm$ 0.027  & 0.482 $\pm$ 0.050  \\
&$\chi^{2}_{dof}$ & 3.8 & 3.1  & 3.4  & 4.1 & 2.4 & 1.5  & 2.6 \\
&$d.o.f.$ & 44  & 44  & 44  & 44  & 44  & 44  & 44 \\
\hline
\end{tabular}
\end{center}

\end{table*}

\begin{table}
\footnotesize
%%\tiny
\caption {The phase-resolved spectral analysis results of PSR
B0540-69} \scriptsize{}\label{table:B0540phrespec}
\medskip
\begin{center}
\begin{tabular}{c c c c c c c}
\hline \hline

& Pulse phase range & Normalization($10^{-3}$)  & Spectral Index  & $\chi^{2}_{dof}$ & d.o.f. \\
\hline

& 0.00-0.16 & 0.50$\pm$0.08 & 1.925$\pm$0.083 & 1.2 & 41 \\
& 0.16-0.20 & 1.84$\pm$0.33 & 2.059$\pm$0.084 & 0.8 & 58 \\
& 0.20-0.24 & 2.59$\pm$0.30 & 2.002$\pm$0.048 & 1.1 & 58 \\
& 0.24-0.28 & 3.70$\pm$0.30 & 2.006$\pm$0.038 & 1.4 & 58 \\
& 0.28-0.32 & 4.60$\pm$0.30 & 1.983$\pm$0.030 & 1.5 & 58 \\
& 0.32-0.36 & 4.40$\pm$0.28 & 1.941$\pm$0.026 & 1.3 & 58 \\
& 0.36-0.40 & 4.45$\pm$0.29 & 1.963$\pm$0.030 & 1.0 & 58 \\
& 0.40-0.44 & 4.02$\pm$0.26 & 1.927$\pm$0.028 & 1.1 & 58 \\
& 0.44-0.48 & 4.20$\pm$0.28 & 1.943$\pm$0.028 & 1.3 & 58 \\
& 0.48-0.52 & 3.92$\pm$0.26 & 1.909$\pm$0.027 & 0.9 & 58 \\
& 0.52-0.56 & 4.22$\pm$0.29 & 1.973$\pm$0.032 & 0.9 & 58 \\
& 0.56-0.60 & 3.72$\pm$0.32 & 2.029$\pm$0.040 & 1.2 & 58 \\
& 0.60-0.64 & 2.21$\pm$0.28 & 1.971$\pm$0.053 & 1.3 & 58 \\
& 0.64-0.80 & 0.86$\pm$0.17 & 2.134$\pm$0.105 & 1.5 & 37 \\

\hline
\end{tabular}
\end{center}

\end{table}

\end{document}